\documentclass[12pt,a4paper,dvips]{article}
\usepackage{epsfig}

\renewcommand{\topfraction}{0.99}
\renewcommand{\bottomfraction}{0.99}
\renewcommand{\textfraction}{0.01}
\begin{document}

\title{\bf Calculation of atmospheric neutrino flux}

\author{\bf{V.Plyaskin} \\
\it{Institute of Theoretical and Experimental Physics,}\\
\it{Moscow,Russia }\\}

\date{March 21, 2001.}

\maketitle

\begin{abstract}
A calculation of the fluxes of primary particles
arriving to the Earth's vicinity as well as those produced in the interactions of  
the primaries with the atmosphere is presented. 
The result of calculations is compared with the experimental data
obtained with the Alpha Magnetic Spectrometer (AMS).
A good agreement of calculated and measured fluxes of charged particles
supports the viability of the atmospheric neutrino flux calculation. 

\end{abstract}

\section{\bf Introduction}
The interpretation of the data on the atmospheric neutrinos obtained
by different experiments \cite{bib-SKam,bib-Kam,bib-IMB,bib-Soud,bib-MAC}
strongly relies on the calculations of neutrino fluxes 
\cite{bib-Hon1,bib-Hon,bib-Barr,bib-Barr1,bib-Vol,bib-Bug,bib-But,bib-Lip1,bib-Batt,bib-LipGeo}.
Starting from a common assumption that the primary cosmic rays arrive 
in the vicinity of the Earth isotropically the calculations differ in the level
of sophistication used  by different authors to describe the production of the
secondaries in the hadronic interactions of primary cosmic particles (mostly 
protons) with the atmosphere and in the way the secondaries are propagated in 
the Earth's magnetic field. Different calculations producing considerably 
different results, the validity of neutrino flux calculations can only be checked
by comparing the predictions of these calculations with the experimental data.
Although the difference in the measured and calculated atmospheric fluxes of 
upgoing and downgoing neutrinos is considered as a clear evidence for 
neutrino oscillations \cite{bib-LipGeo} it is nevertheless 
recognized \cite{bib-LipEW2}, that the same calculations \cite{bib-Hon,bib-LipEW1}
seem to fail to properly describe the measured east-west asymmetry of atmospheric neutrino 
fluxes \cite{bib-SKamEW} 
related to a well known  \cite{bib-John,bib-Alvar} observation that due to the 
influence of the Earth's (nearly dipole) magnetic field the fluxes of the primary
predominantly positive (p,$\alpha$-particles) hadronically interacting 
cosmic rays reaching the atmosphere contain more particles coming from the west
than from the east. Thus, more accurate and detailed 
calculations describing propagation of primary cosmic rays and their interactions
with the atmosphere is needed. 

The recent data from the Alpha Magnetic Spectrometer (AMS) \cite{bib-AMS0,bib-AMS1,bib-AMS2}
provide a vast amount of information on the fluxes of both primary and 
secondary particles detected in the near Earth orbit covering most of the Earth's 
surface. The AMS data display not only the common features of the magnetic cutoff 
dependence on the distance from the magnetic equator but also reveal the presence 
of particles well under the cutoff \cite{bib-AMS1,bib-AMS2}. These particles originate
from the production and eventual decay of the secondaries produced in the inelastic 
interactions of the primary cosmic particles with the atmosphere and consequently carry 
the information about neutrinos produced in decays of the secondaries.
A comparison of the results obtained in the calculations with the fluxes of the 
secondaries measured with AMS represents a means to prove or disprove the validity 
of the results of the calculations concerning the predictions of atmospheric 
neutrino fluxes. 

In the following the result of calculation of the near Earth fluxes of charged particles 
and atmospheric neutrinos is presented. First, the model used and its parameters is 
described. Section 3 describes the origin of the particles near the Earth. Section 4 is 
devoted to results of calculations of 
fluxes of charged particles at the altitude of 400 km from the Earth's surface 
corresponding to the AMS flight altitude. The simulated fluxes of charged particles are compared 
with those measured with AMS in section 5.  
Section 6 is devoted to the results of calculations of atmospheric neutrino fluxes 
at the ground level in general. The neutrino fluxes at different experimental sites are discussed 
in section 7. 
The last section gives a summary of the results and draws some conclusions.  

\section{\bf Description of the model}

The calculation uses the Monte Carlo technique and is based on the GEANT3/GHEISHA package 
\cite{bib-GEANT}. This package is extensively used for the simulation of processes in various
detectors of high energy physics experiments.  To use this package for the simulation on a 
scale of propagation of cosmic rays in the Earth's magnetic field, their interactions with
the Earth's atmosphere and the Earth itself, followed by production, propagation and decay
of secondary particles, the parameters of the corresponding media are adapted to 
scale them from a thousands of kilometer scale of cosmic ray travel and interaction to
a centimeter scale characteristic for GEANT based simulations.

The Earth is modeled to be a sphere of 6378.14 km radius of a uniform density of 
5.22 g/$\rm cm^3$ derived from the total mass of the Earth. The atomic and nuclear properties
of the Earth are taken to be those of Ge - the closest in density (5.32 g/$\rm cm^3$) element.
The Earth's atmosphere is modeled by 1 km thick variable density layers of air extending up
to 71 km from the Earth's surface. The density change with altitude is calculated using a Standard
Atmosphere Calculator \cite{bib-atm}. 

The Earth's magnetic field for the year 1998 is calculated 
according to World Magnetic Field Model (WMM-2000) \cite{bib-WMM} with 6 degrees of spherical 
harmonics. The WMM-2000 is a predictive model based on the magnetic field measurements on 
a global scale.
 
The flux ($\Phi$) of high energy primary protons in the Solar system is parametrized 
on the basis of the existing experimental data 
\cite{bib-AMS1,bib-p1,bib-p2,bib-p3,bib-p4,bib-p5} by a power low in 
rigidity (R), $\Phi(R) \sim R^{-\gamma}$ with $\gamma$=2.79.
At the energies below several GeV the spectrum is corrected for the solar activity to be similar 
to that of the year 1976 \cite{bib-p6} - the year of a similar solar activity as in the year 
1998 according to the 11 year solar cycle.
Primary protons are isotropically emitted from a sphere of 10 Earth's radii. The long 
distance from the Earth of the emission point is sufficient for a particle to endure sufficient 
influence of the Earth's magnetic field to form an adequate topology of the flux in the Earth's vicinity.

Secondary particles produced in the interactions with the atmosphere and the Earth are traced in the
magnetic field following a normal GEANT interaction/decay procedure.

\section{\bf Origin of particles near Earth}

\begin{figure}[hp]
\begin{center}
\mbox{\epsfig{file=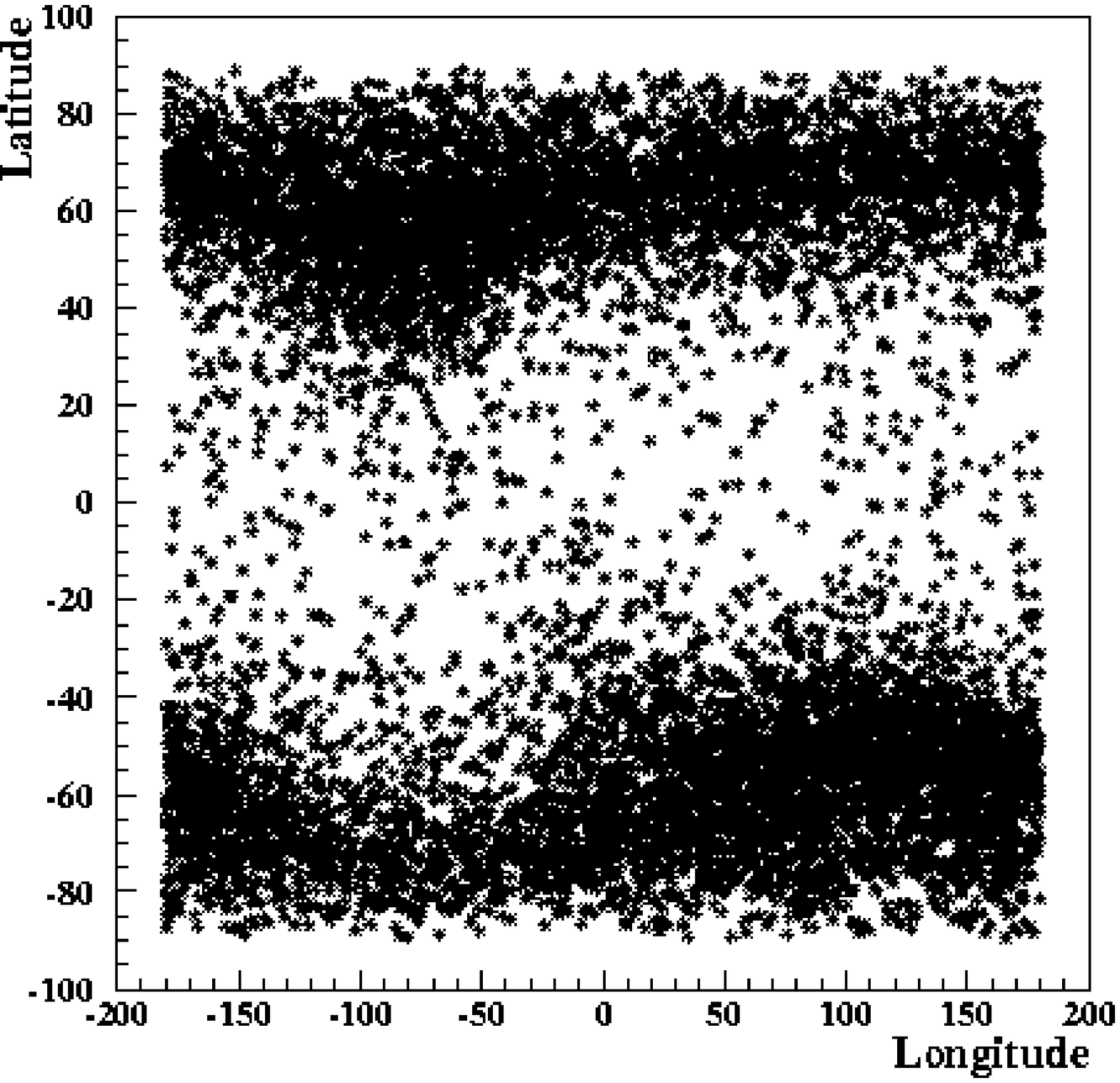,height=10 cm,width=10cm}}
\mbox{\epsfig{file=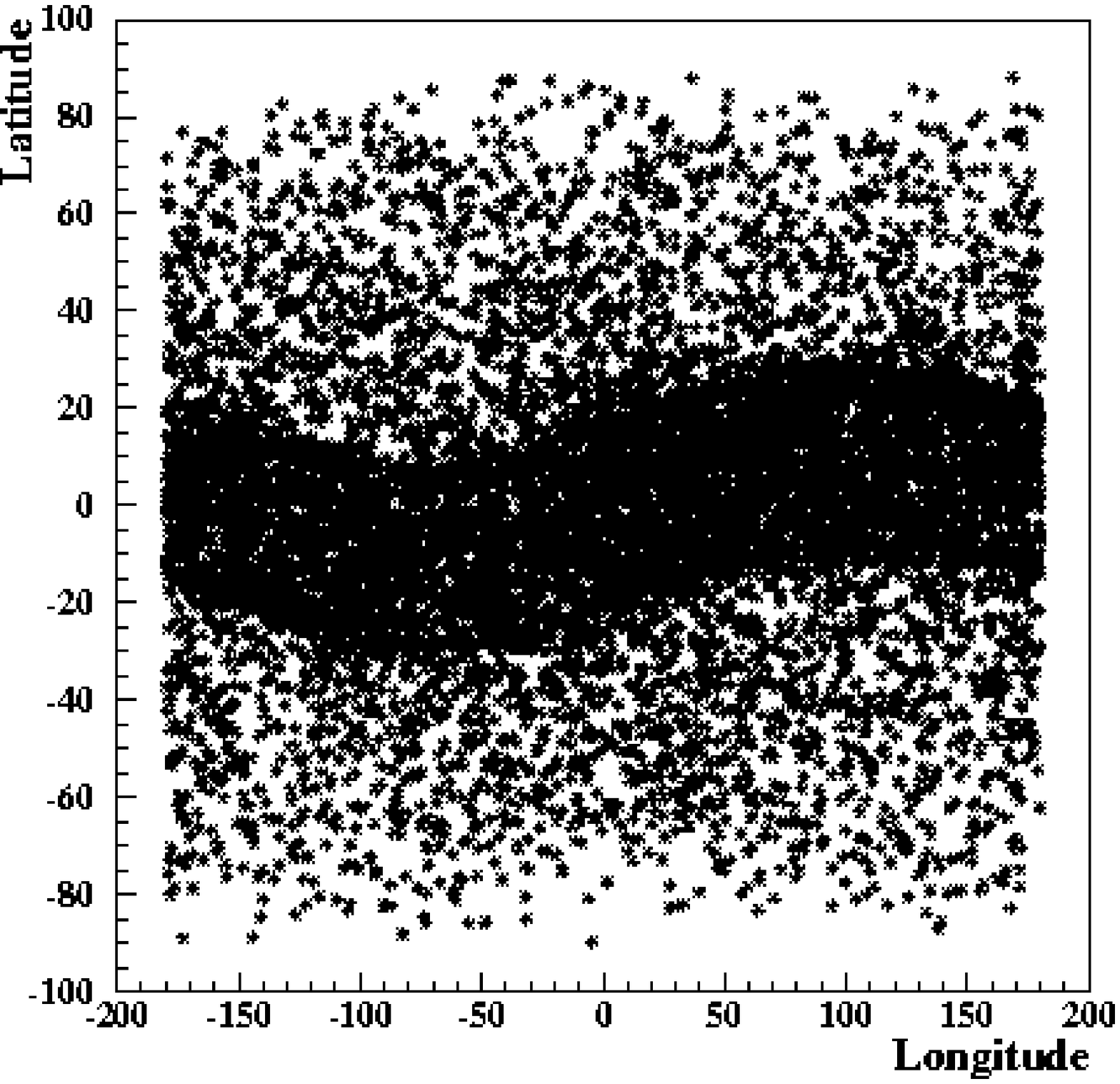,height=10 cm,width=10cm}}
\caption{%
Top: The points of impact of primary cosmic rays on the Earth.
Bottom: The points of origin on the 10 Earth's radii sphere of primary 
cosmic rays reaching the Earth.}
\label{fig:pole}
\end{center}
\end{figure}

The shape of magnetic field determines the behaviour
of charged particles at their approach to the Earth and the eventual motion of primaries and secondaries
in the Earth's vicinity.
The channeling of the particles in the Earth's magnetic field leads to a well known effect that the
flux of particles coming from space is at its maximum at the magnetic poles (see top Fig.\ref{fig:pole}).
This effect is associated with the magnetic cutoff of low momenta particles in the equator region.
A peculiar feature of particle channeling from the outer space is that most of the particles detected in
the pole region in the Earth's vicinity have their origin in the equator region of the isotropically
shining sphere at 10 Earth's radii used in the simulation (bottom Fig.\ref{fig:pole}).
The efficiency of channeling of the particles depends on the emission parameters such as particle 
momentum and emission direction. 

\begin{figure}[hp]
\begin{center}
\mbox{\epsfig{file=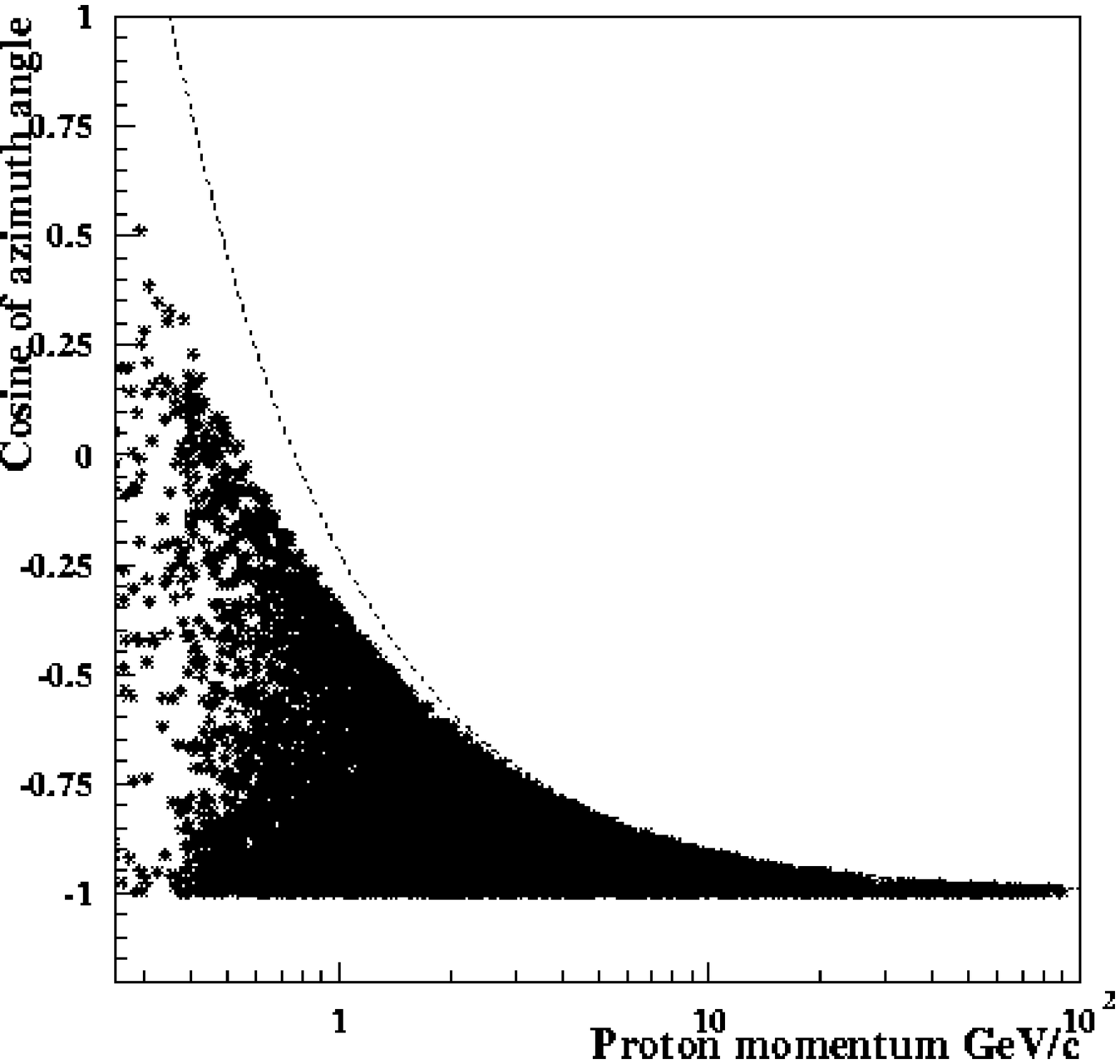,height=10 cm,width=10cm}}
\mbox{\epsfig{file=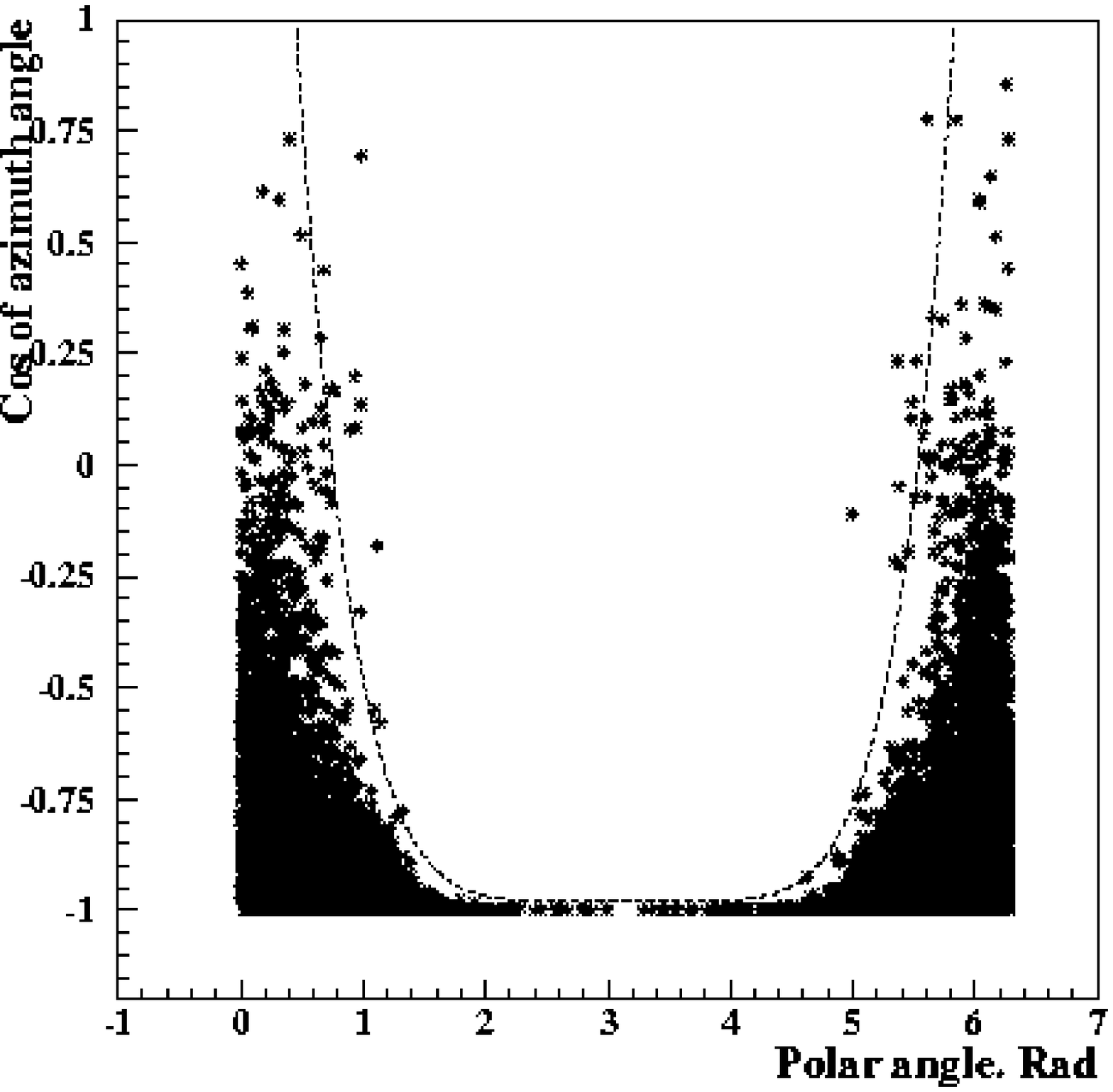,height=10 cm,width=10cm}}
\caption{%
The phase space in cos($\rm \Theta_{emission}$)/momentum (top) and 
cos($\rm \Theta_{emission}$)/$\phi_{emission}$ (bottom) planes of particles reaching 
the Earth's vicinity. 
The dashed lines show the limits used in the simulation.} 
\label{fig:ptheta}
\end{center}
\end{figure}

Figure \ref{fig:ptheta} 
shows the emission parameters of the particles reaching the Earth. One can see in figure \ref{fig:ptheta}
that small momentum particles even emitted in the direction opposite to the Earth can reach the
Earth's vicinity whereas high momentum particles can do that only when the initial direction is 
close to a straight line between the emission point and the Earth. The dashed lines in 
figure \ref{fig:ptheta} show the cuts on the emission parameters used in the
simulation.
An additional cut is set on the momentum of primary protons. It is set to be of 0.5 GeV/c
for two reasons. First, due to the detector material thickness the minimum kinetic energy 
required for a proton to trigger AMS is $\sim$ 0.13 GeV corresponding to a proton
momentum of 0.5 GeV/c. Secondly, the kinetic energy threshold of charged pion production 
in p-nucleus collisions is $\sim$ 0.16 GeV \cite{bib-pN} corresponding to a proton momentum
of 0.54 GeV/c. Protons with energies below the threshold can not produce charged pions which 
represent the source of detectable secondaries and eventually of atmospheric neutrinos. 
The cut on the proton momentum eliminates the few events in the region forbidden by the cut 
of the bottom Fig.\ref{fig:ptheta}. 
  
\newpage

\newpage

\newpage

Of all protons with the momentum spectrum described above uniformly 
emitted from the sphere of 10 Earth's radii only a $7.4 * 10^{-4}$ part reaches the Earth's atmosphere.
In the subsequent simulation only particles with initial kinematics compatible with those 
reaching the Earth's atmosphere are traced the whole way until they interact with the atmosphere or
the Earth itself or/and continue the journey beyond a sphere of 15 Earth's radii. 
The charged secondaries from the interactions are traced until they go below a limit 
of 0.125 GeV/c.

\begin{figure}[htb]
\begin{center}\mbox{
\epsfig{file=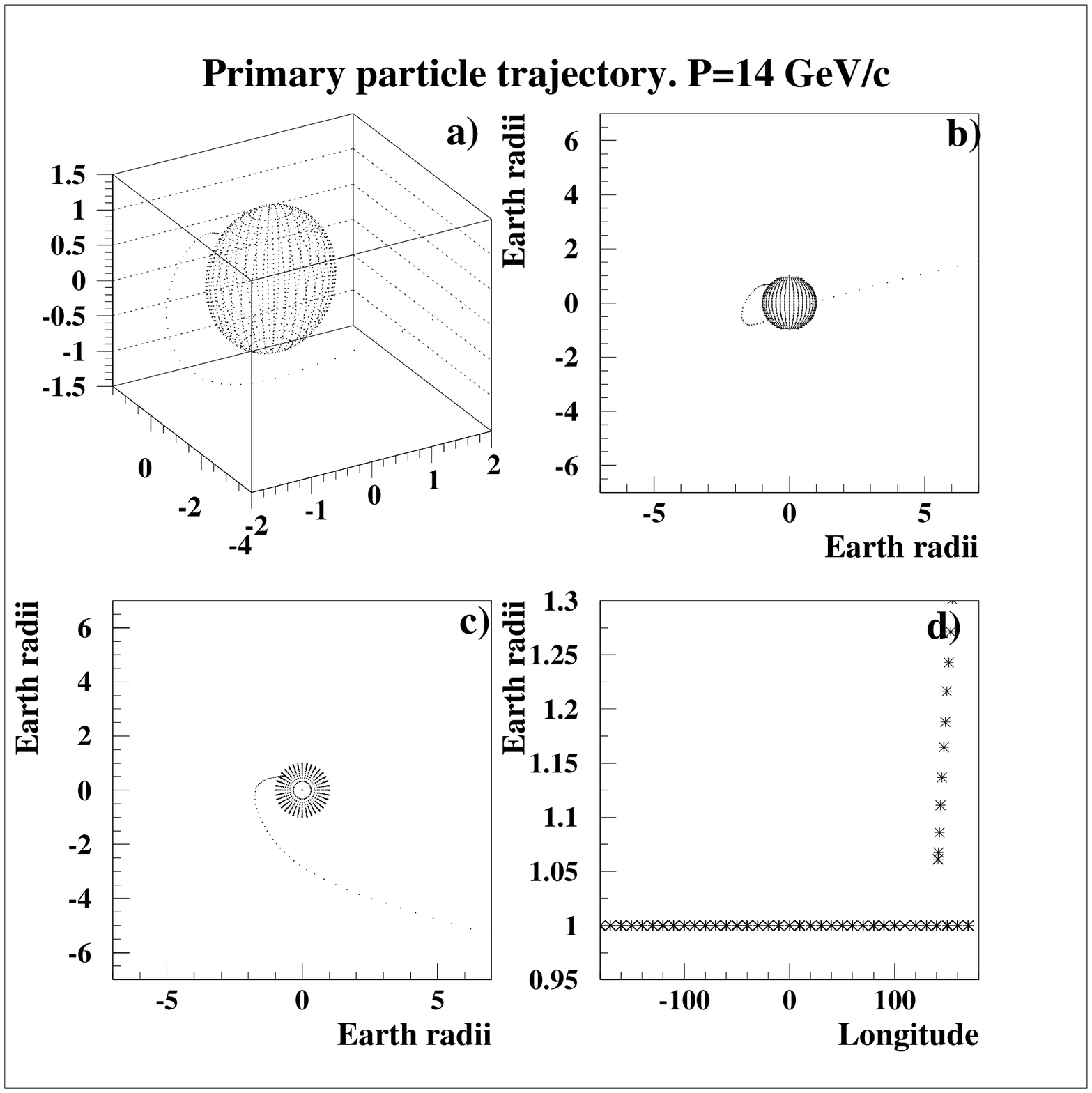,height=15 cm,width=15cm}}
\caption{%
Primary proton trajectory in the Earth's magnetic field for a 
particle reaching the Earth. 
a) General view, b) View from 90 degree west c) View from the north pole,
d) View at the Earth approach.}
\label{fig:pp14}
\end{center}
\end{figure}

Figure \ref{fig:pp14} shows an example of the trajectory of a particle reaching the Earth. (Note that
the AMS flight altitude of 400 km corresponds to 1.06 Earth's radii). 
Due to influence of the Earth's magnetic field the direction of a particle at 
its entrance to the atmosphere is by far different from the initial direction. 

\begin{figure}[htb]
\begin{center}\mbox{
\epsfig{file=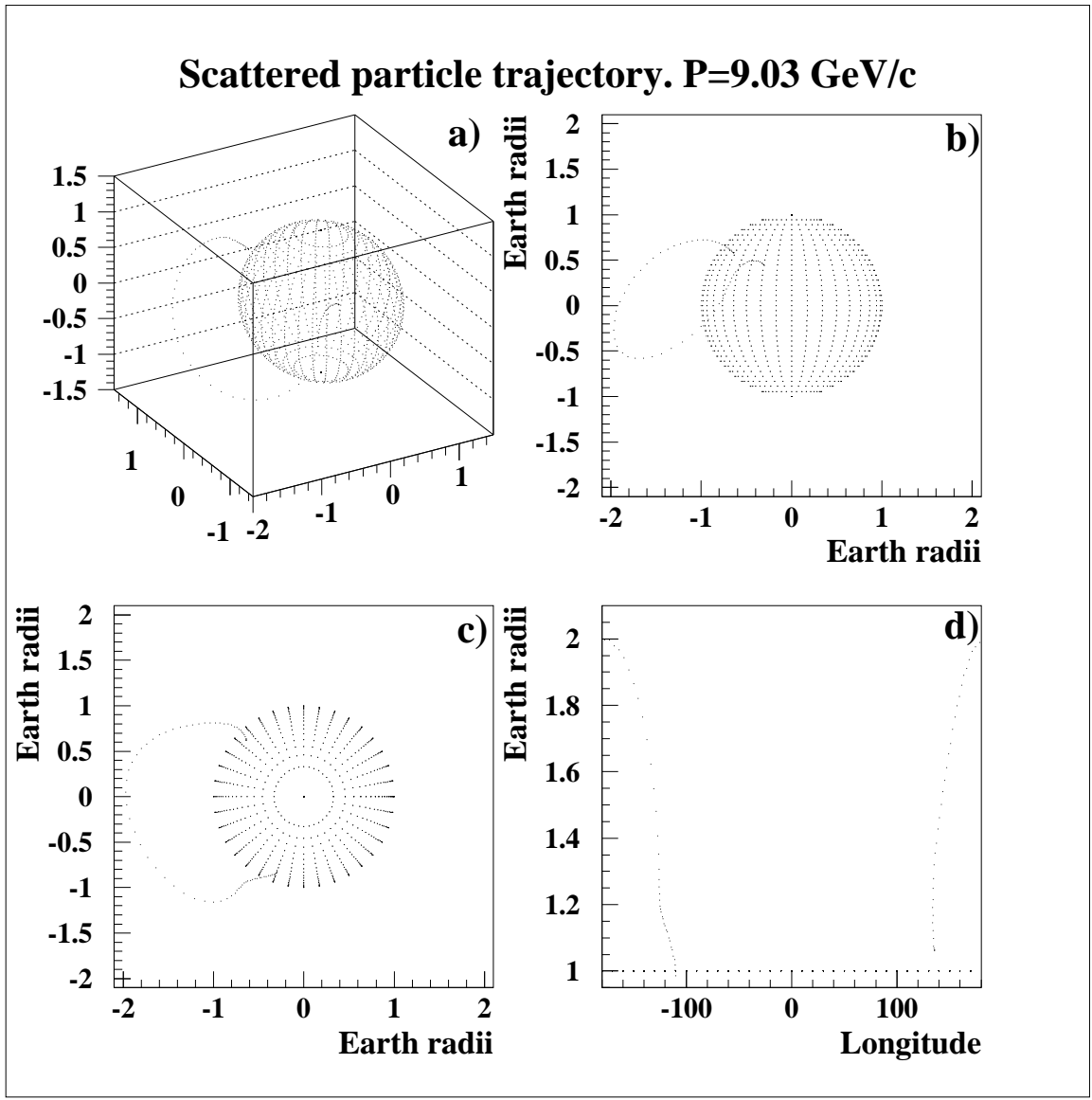,height=15 cm,width=15cm}}
\caption{%
Trajectory of proton scattered in the atmosphere.
(a) General view. (b) View from 90 degree west. (c) View from the north pole.
(d) View of the scattering point and the Earth approach.}
\label{fig:prs9_03}
\end{center}
\end{figure}

Primary cosmic particles may endure an elastic or inelastic scattering in the atmosphere 
thus changing their direction and/or momentum. Figure \ref{fig:prs9_03} shows the trajectory of 
such a particle which eventually reenter the atmosphere but at a place far away from the point of 
initial interaction. As it will be shown later, there is a considerable amount of such particles in 
the total flux of incoming cosmic rays. 

\begin{figure}[htb]
\begin{center}\mbox{
\epsfig{file=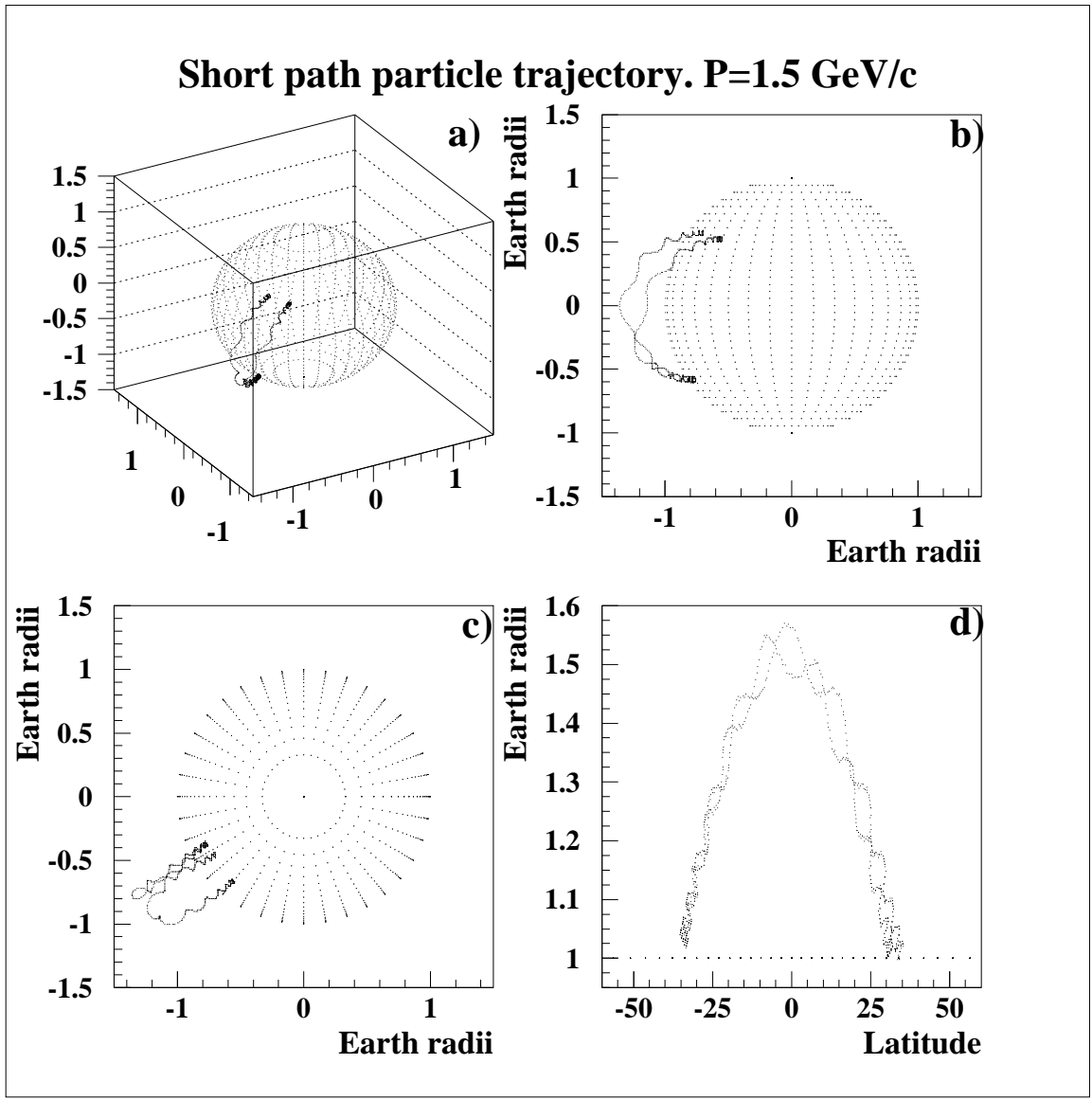,height=15 cm,width=15cm}}
\caption{%
Trajectory of secondary positive particle with short path length.
(a) General view. (b) View from 90 degree west. (c) View from the north pole.
(d) Altitude at different latitude.}
\label{fig:psp1_5}
\end{center}
\end{figure}

Secondary particles produced in the inelastic interactions of primary 
cosmic rays with the atmosphere follow a usual pattern of hadronic or electromagnetic cascade and 
most of them quickly end up in the atmosphere itself or in the Earth. However, under the influence 
of the Earth's magnetic field some (predominantly stable) particles from the atmospheric showers 
escape upward from the atmosphere and travel long distances before being absorbed in the atmosphere.
The trajectory of such particle depends on the kinematics of the particle production and the 
geomagnetic coordinates of the origin. Unless they interact with the atmosphere the particles
with the same rigidity (p/z) follow the same trajectory. Figure \ref{fig:psp1_5} shows the trajectory
of a positive particle produced in the atmosphere in the northern hemisphere, reflected at a mirror 
point in the southern hemisphere before being absorbed at an approximatively the same latitude but
a different longitude. 

\begin{figure}[htb]
\begin{center}\mbox{
\epsfig{file=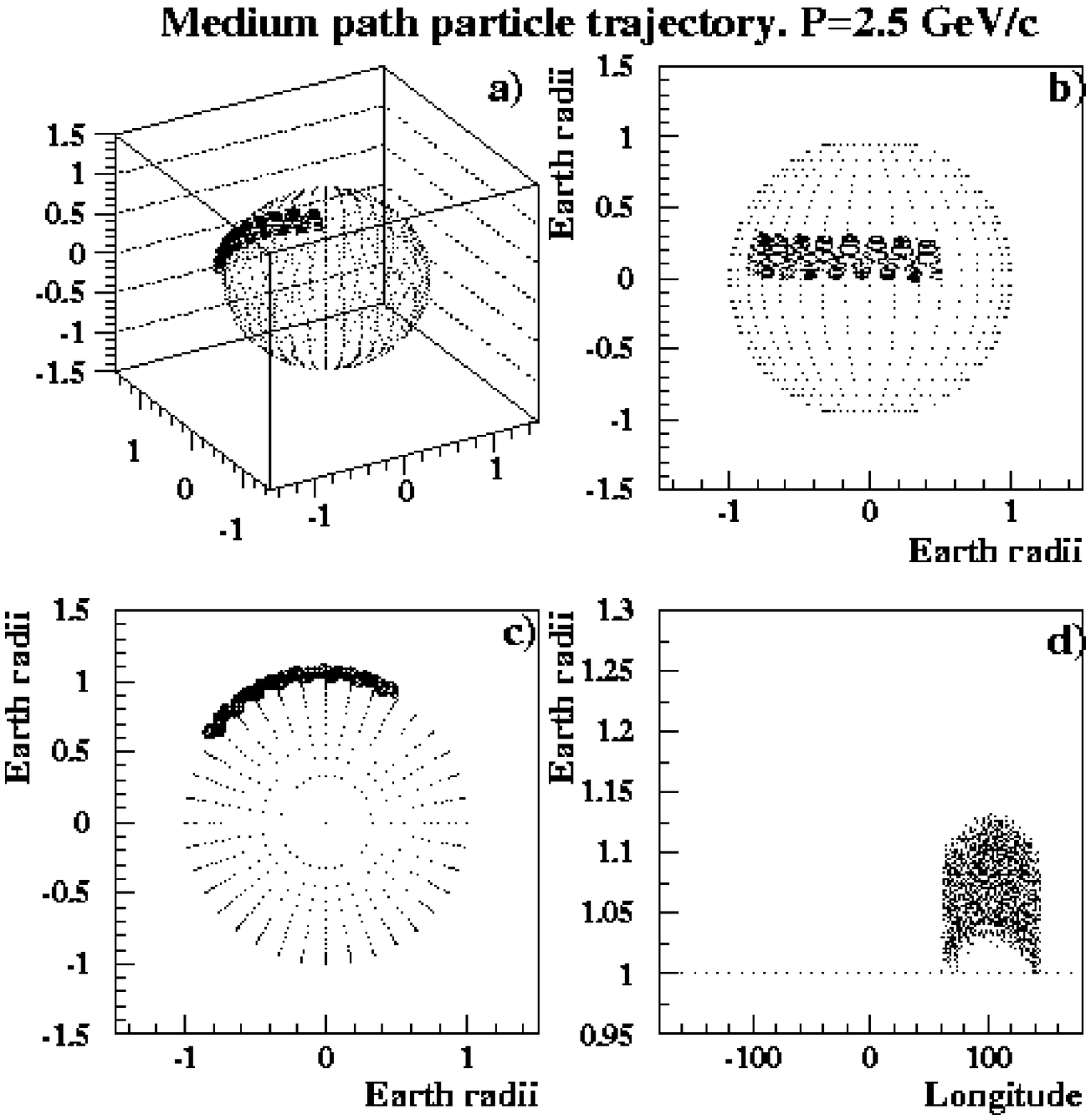,height=15 cm,width=15cm}}
\caption{%
Trajectory of multi bouncing secondary particle with medium path length.
(a) General view. (b) View from 90 degree west. (c) View from the north pole.
(d) Altitude at different longitude.}
\label{fig:pmp2_5}
\end{center}
\end{figure}

\begin{figure}[htb]
\begin{center}\mbox{
\epsfig{file=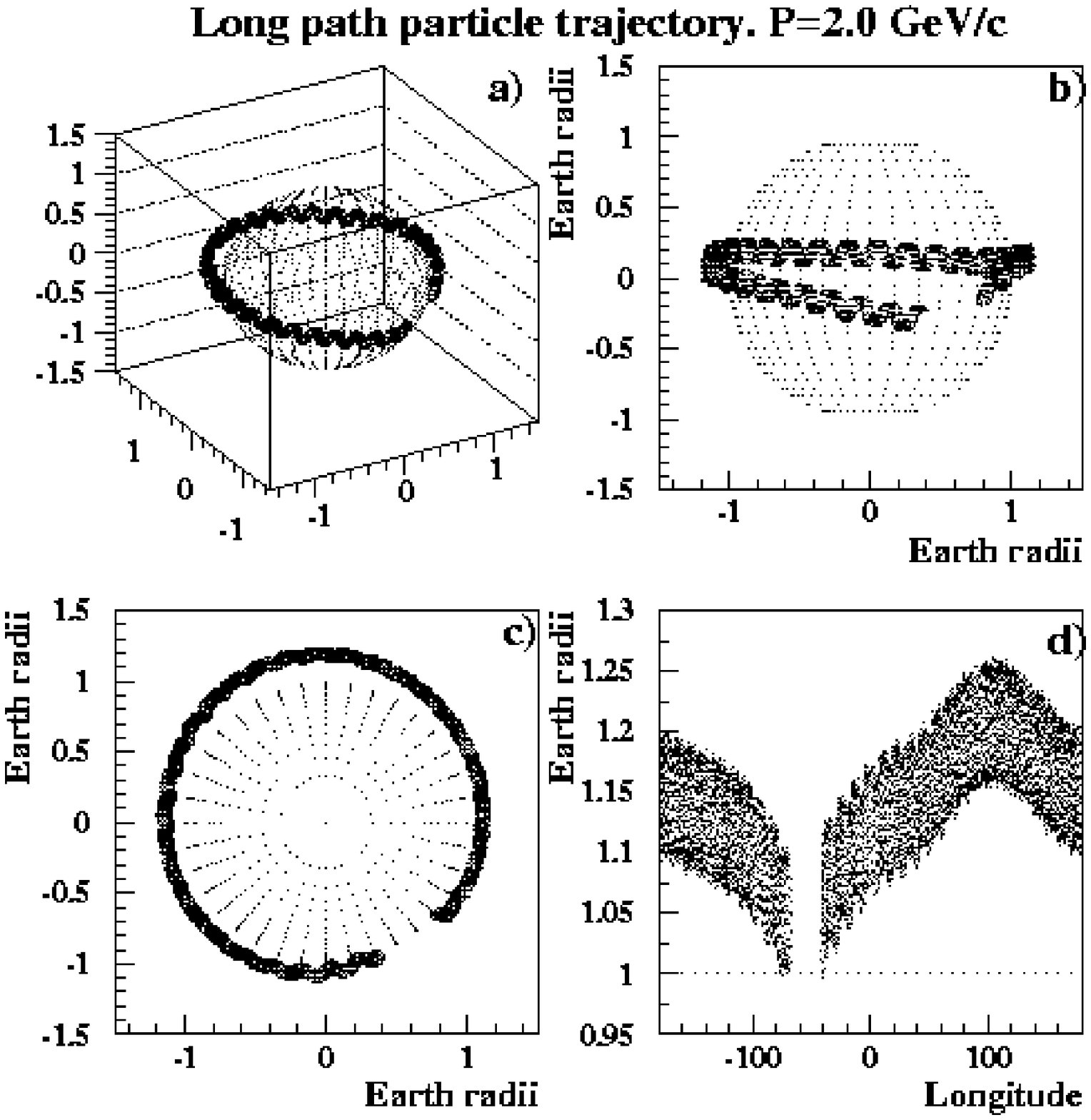,height=15 cm,width=15cm}}
\caption{%
Trajectory of multi bouncing secondary particle with long path length.
(a) General view. (b) View from 90 degree west. (c) View from the north pole.
(d) Altitude at different longitude.}
\label{fig:plf2_0}
\end{center}
\end{figure}

\begin{figure}[htb]
\begin{center}\mbox{
\epsfig{file=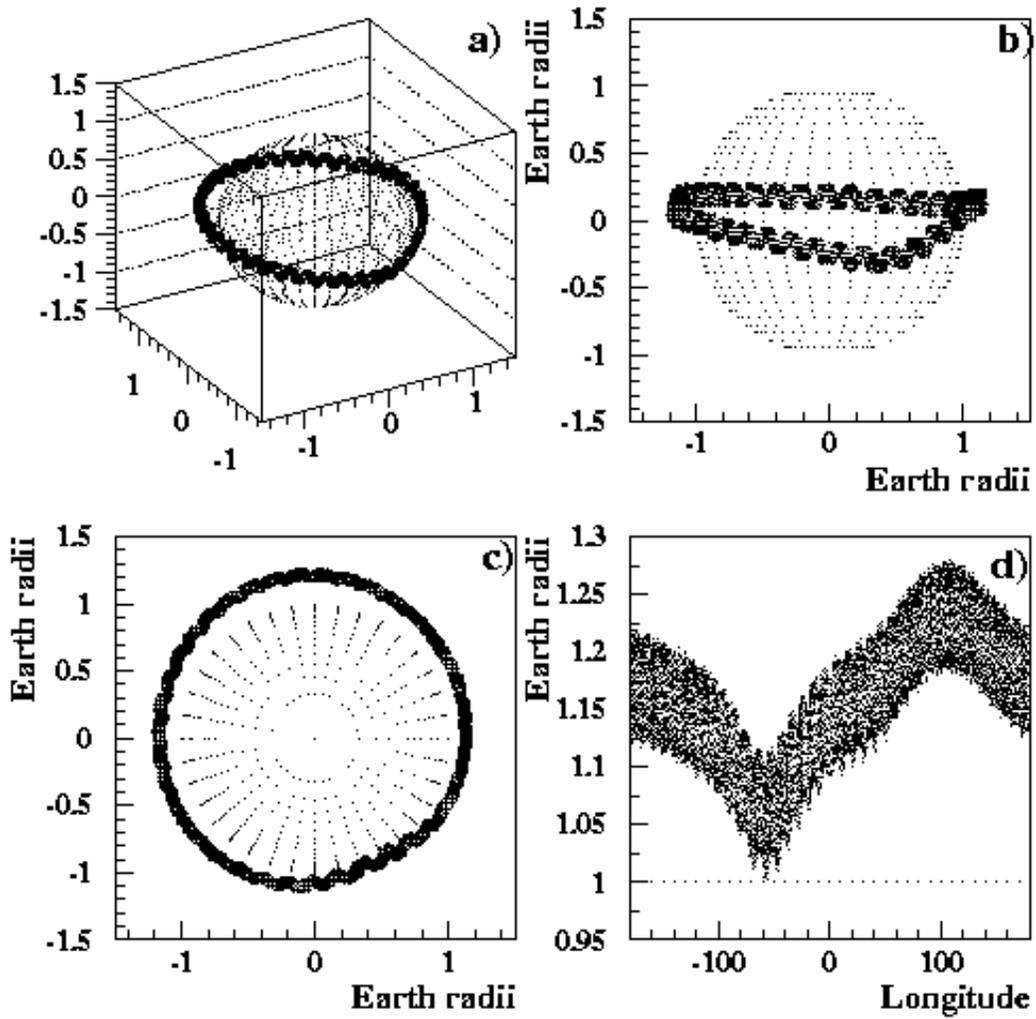,height=15 cm,width=15cm}}
\caption{%
Closed orbit trajectory of multi bouncing secondary particle.
(a) General view. (b) View from 90 degree west. (c) View from the north pole.
(d) Altitude at different longitude.}
\label{fig:ptr1_66}
\end{center}
\end{figure}

The spiraling of the particles, their motion from and to the Earth's surface 
as well as their bouncing across and drifting along the magnetic equator is determined by the
shape of the Earth's magnetic field. 
Depending on the production place and direction with respect to the Earth's surface secondary 
particles can make a return trip as illustrated in Fig.\ref{fig:psp1_5}, a one way trip across the 
magnetic equator, a multiple bounce (Fig.\ref{fig:pmp2_5}) or even a round the globe trip 
(Fig.\ref{fig:plf2_0}). At the limit a secondary particle produced in the atmosphere over a certain 
limited region of the Earth (called South Atlantic Anomaly) can have a closed trajectory 
(Fig.\ref{fig:ptr1_66}) and could live forever if it would not interact with the atmosphere at 
the place where it was created nor lose its energy on synchrotron radiation.

As one can see in figures \ref{fig:psp1_5}d, \ref{fig:pmp2_5}d, \ref{fig:plf2_0}d and \ref{fig:ptr1_66}d
the particles in their motion pass the level of the AMS flight (1.06 Earth's radii) going up and down 
several times.
The geographic coordinates of the place over which a secondary particle can be detected at a certain
altitude is connected with the place of origin of the particle.

\newpage

\newpage

\newpage

\newpage

\newpage

\newpage

\newpage

 One of the consequences of the fact 
that in its motion the same particle can cross a detector several times is that the measured flux of
such upgoing and downgoing particles is the same. A build up of the measured rate, especially in the
region close to the magnetic equator, connected with the multiple crossing of the detector plane 
rate, does not produce  secondary particles unless the spiraling particle
passes through and interacts with the atmosphere. The motion of the particle being determined by
the Earth's magnetic field, negative particles travel along the same trajectories as the positive 
ones but in the opposite direction.

The production of secondary pions and protons resulting from the interactions of cosmic (mostly
He) nuclei with the atmosphere is treated using the superposition approximation \cite{bib-Schu}.
 
The calculation presented below takes into account all effects connected with production, propagation,
interaction and decay of all particles of the atmospheric shower.                
   
\section{\bf Particle flux}  
The results of Monte Carlo simulation presented in the following correspond to a statistics of
$\sim 1.5*10^{10}$ protons isotropically emitted from a sphere of 10 Earth's radii. This is equivalent
to a 0.62 ps exposure. 

\begin{figure}[htb]
\begin{center}\mbox{
\epsfig{file=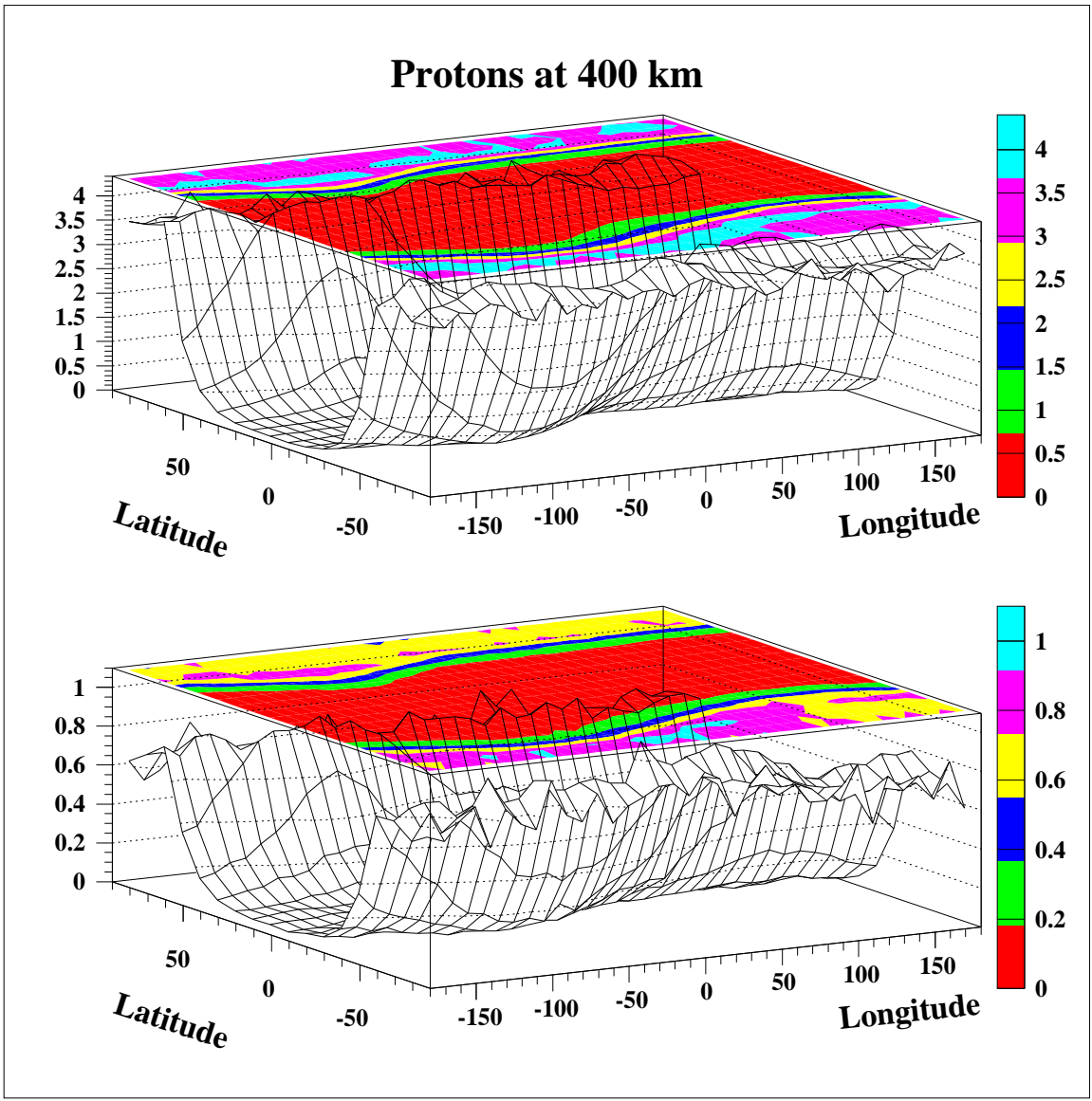,height=15 cm,width=15cm}}
\caption{%
Proton rate (in KHz/$\rm m^2~sr$) at 400 km altitude averaged over all directions.
Top: earthward protons, bottom: spaceward protons.}
\label{fig:501}
\end{center}
\end{figure}

The dependence on the geographical coordinates of the rate of protons crossing a sphere surrounding 
the Earth at the altitude of 400 km is shown in Figure \ref{fig:501} for downward (top picture) and
upward (bottom picture) protons. The rate includes all i.e. primary and secondary, protons and is 
averaged over all directions in the corresponding hemispheres.

A typical picture with high rate near the magnetic poles and the depleted by the magnetic cutoff 
regions near the magnetic equator is clearly seen in the downward proton rate behaviour. 
The geographical pattern of upgoing protons is similar in shape but at much lower particle rate.
          
\begin{figure}[hp]
\begin{center}\mbox{
\epsfig{file=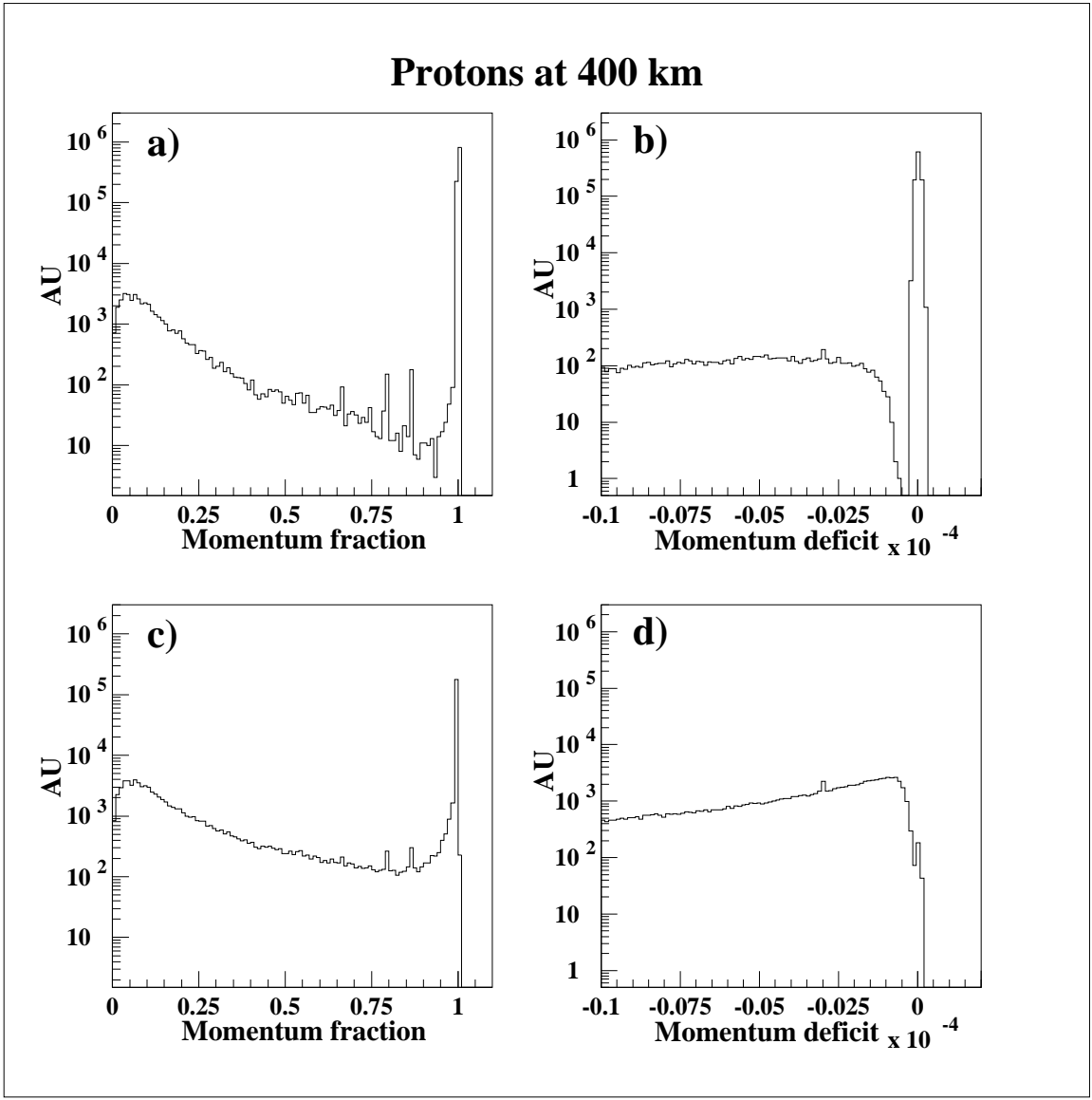,height=15 cm,width=15cm}}
\caption{%
Partial (with respect to initial proton) momentum of protons detected at 400 km. 
(a) All downgoing protons. (b) Close up on the close to unity region of (a).
(c) All upgoing protons. (d) Close up on the close to unity region of (c).}
\label{fig:504}
\end{center}
\end{figure}

The flux of protons going toward the Earth (Figs.\ref{fig:504}a and b) consists of original
protons from space i.e. those with $p/p_{init}>$0.9999996, where $p_{init}$ is the initial cosmic
proton momentum and $p$ is the momentum of the proton crossing the detection sphere
(see Fig.\ref{fig:pp14}), protons scattered in the atmosphere (0.995$<p/p_{init}<$0.9999996.
Fig.\ref{fig:prs9_03}) and secondary protons ($p/p_{init}<$0.995. 
Figs.\ref{fig:psp1_5},\ref{fig:pmp2_5},\ref{fig:plf2_0},\ref{fig:ptr1_66}). 
The flux of upgoing protons (Figs.\ref{fig:504}c and d) contains reflected protons most of which have
lost some energy in the atmosphere as well as scattered protons on their way 
from the Earth and the secondary protons. 

\newpage

\newpage

\newpage

\begin{figure}[h]
\begin{center}\mbox{
\epsfig{file=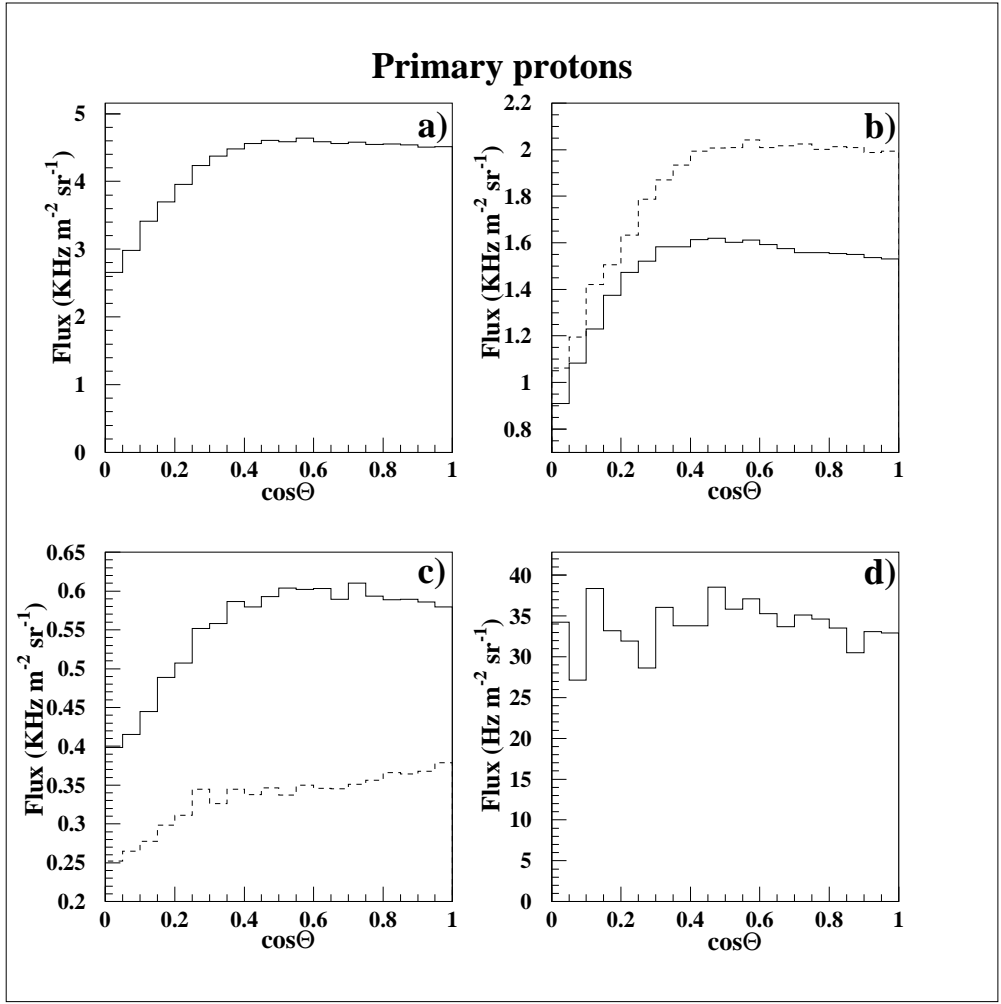,height=13 cm,width=13cm}}
\caption{%
Dependence of primary proton flux at 400 km on the proton incidence angle with respect to nadir.
(a) All momenta. (b) Solid line:$p<1.5$ GeV/c, dashed line:$1.5<p<5.0$ GeV/c.
(c) Solid line:$5.0<p<10.0$ GeV/c, dashed line:$10.0<p<40.0$ GeV/c.
(d) p$>40.$ GeV/c.} 
\label{fig:505o}
\end{center}
\end{figure}

The important feature of the primary flux relevant to neutrino production is that up to momenta as high 
as $\sim$40 GeV it becomes anisotropic
for directions close to parallel to the Earth's surface.
(see Fig.\ref{fig:505o}). This decrease is not in contradiction
with the AMS measurement because the orientation of the AMS detector during the flight did not allow to
measure the horizontal flux.   

\newpage

Figure \ref{fig:505} illustrates the behaviour of the flux of primary protons.
Figure \ref{fig:505}a shows the dependence of the cutoff on the magnetic latitude.
The cutoff momentum is about 10 GeV/c at the equator and goes down to zero at the poles. 
One can see, however, that there are some under cutoff particles even in the flux of
primary protons. These under cutoff particles are present at all magnetic latitudes 
(Figs.\ref{fig:505} a and d). The other feature of these protons is that they are moving         
predominantly, and in the magnetic equator region exlusivly (Figs.\ref{fig:505}b and c),
toward the Earth.    

The scattered protons (Fig.\ref{fig:506}) are mostly detected in the regions of the flux
maxima close to the magnetic poles (Fig.\ref{fig:506}a). These are mostly albedo protons 
(Figs.\ref{fig:506}b and d) although there is a considerable amount of those 
(Figs.\ref{fig:506}b and c) reentering the atmosphere (see fig.\ref{fig:prs9_03}).

Secondary protons (Fig.\ref{fig:507}) produced at the magnetic poles add up to the albedo flux
(Fig.\ref{fig:507}b) whereas those produced in the regions closer to the equator
follow the trajectories illustrated in Figs.\ref{fig:psp1_5},\ref{fig:pmp2_5},\ref{fig:plf2_0},
\ref{fig:ptr1_66} 
producing equal fluxes of upgoing and downgoing undercutoff particles (Fig.\ref{fig:507}b).
The impact points of particles with multiple crossing of the detection sphere are clearly seen in 
Fig.\ref{fig:507}a. The rate of under cutoff protons in the magnetic equator region is 
dominated by the long traveling protons (see Figs.\ref{fig:pmp2_5},\ref{fig:plf2_0}).

Positrons and electrons resulting from the decays of secondary pions display the behaviour similar 
to that of secondary protons allowing for much faster energy losses in the atmosphere due to
the difference in radiation and nuclear interaction lengths.

\section{\bf Particles detected by zenith facing AMS}

To compare the results of calculations with the AMS measurements \cite{bib-AMS1,bib-AMS2} 
the overall fluxes produced in the simulation were restricted to the AMS acceptance 
during the zenith facing flight period.
\cite{bib-AMS0,bib-AMS3}. In accordance with the AMS construction no particle with kinetic energy
below $\sim 120$ MeV can trigger AMS. 
The AMS orbit spans within the geographic latitude of $\pm 51.7^o$. 
The corresponding cuts are applied on the MC simulated data. The result of Monte Carlo
is an absolute one. No renormalization is applied to make a comparison with the data.  
As in AMS publications the spectra are presented as a function of kinetic energy of the particles. 
Figure \ref{fig:508m} shows the spectra of protons obtained in the simulation for different
positions of AMS with respect to the magnetic equator. The result of simulation correctly
reproduces the characteristic features present in the AMS data (black dots in the figure) such as:

\begin{itemize}

  \item the value of the cutoff at different magnetic latitudes;
  \item the presence of protons with under cutoff momenta at all magnetic latitudes;
  \item the equality of fluxes of under cutoff protons going to and from the Earth;  
\newpage

\begin{figure}[hp]
\begin{center}\mbox{
\epsfig{file=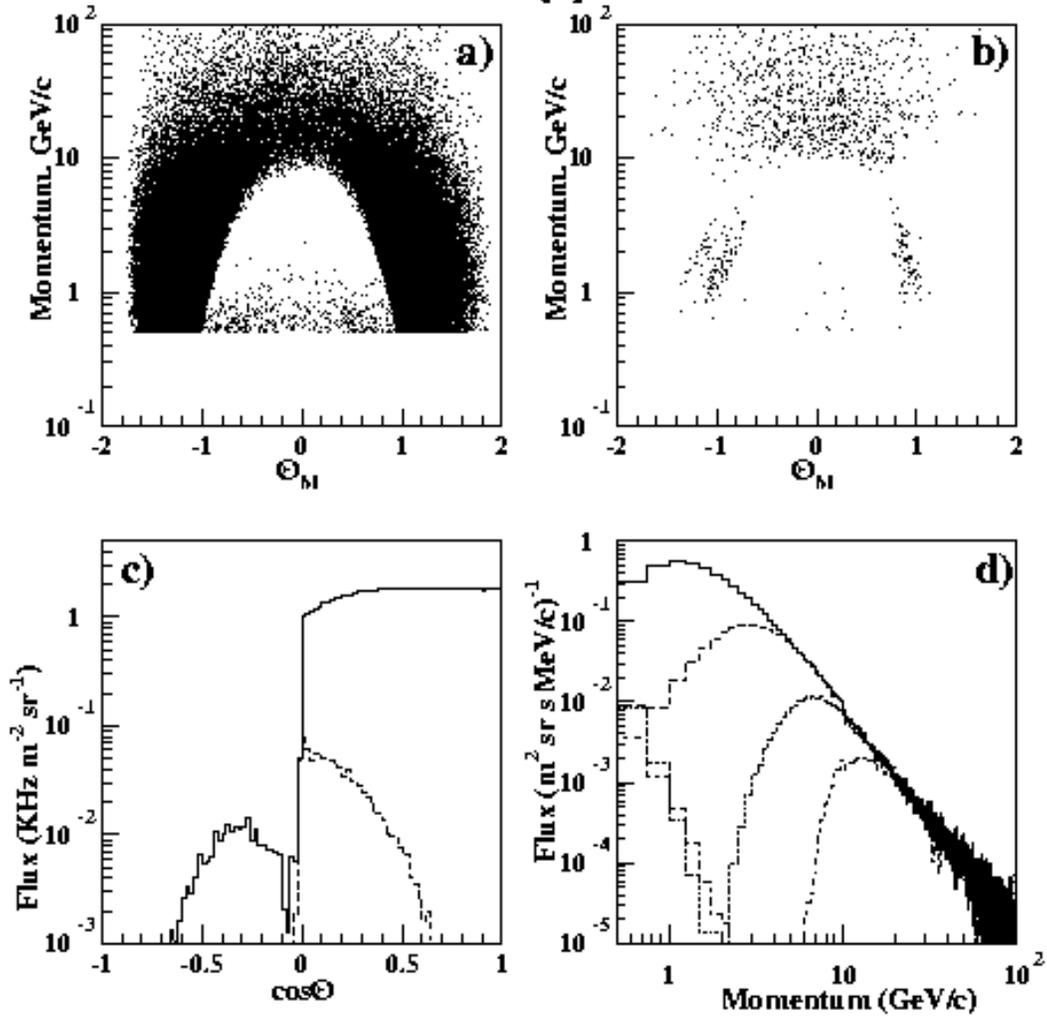,height=15 cm,width=15cm}}
\caption{%
Primary protons detected at 400 km.
Momentum at different magnetic latitude for (a) downgoing and (b) upgoing protons.
(c) Average flux at different angles with respect to nadir. Solid line - all primary protons, dashed line - only 
under cutoff protons in the equator ($|\Theta_M|<$0.9) region.
(d) Momentum spectra at different magnetic latitudes ($\Theta_M$). Solid line - 0.9$<\Theta_M<$1.1,
dashed line  $0.7<\Theta_M<$0.9,dotted line - $0.4<\Theta_M<$0.7, dashed-dotted line -$|\Theta_M|<$0.4}
\label{fig:505}
\end{center}
\end{figure}

\newpage

\begin{figure}[hp]
\begin{center}\mbox{
\epsfig{file=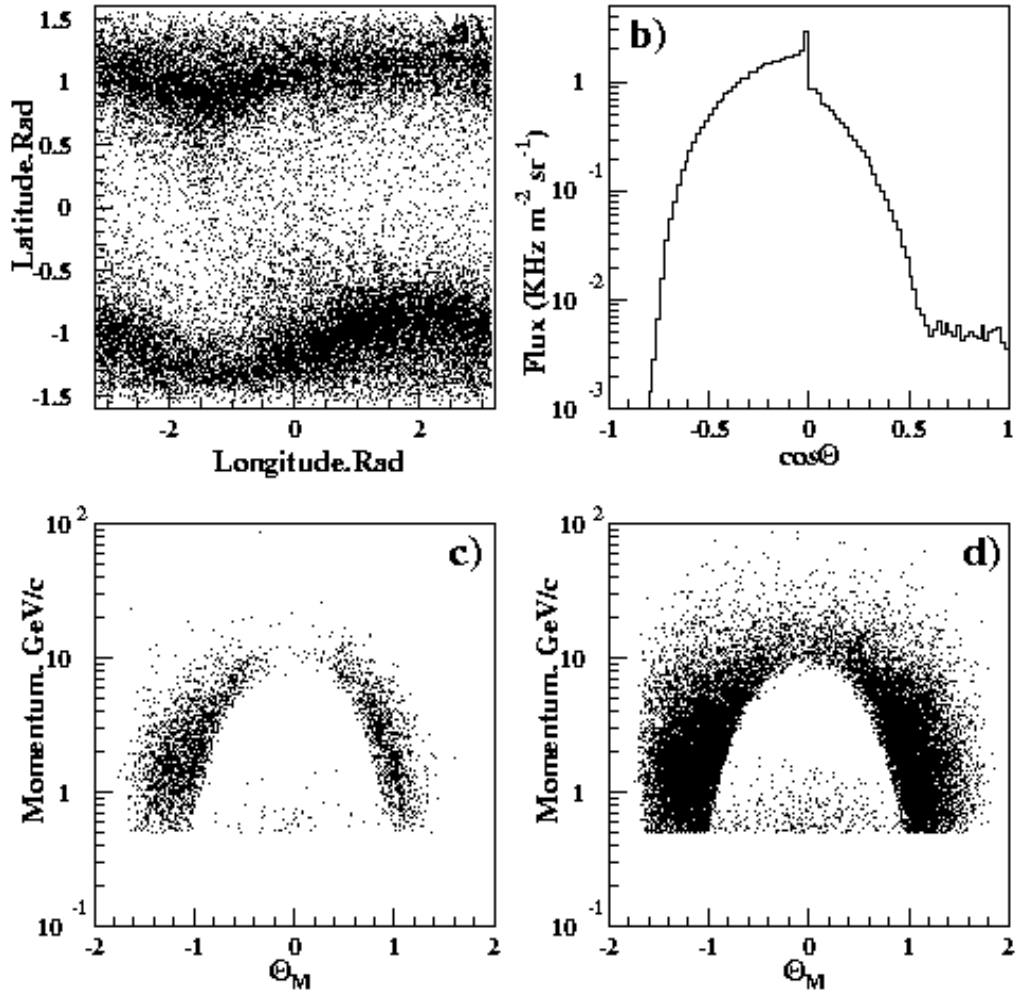,height=15 cm,width=15cm}}
\caption{%
Scattered protons detected at 400 km. 
(a) Geographic coordinates of the crossing points. (b) Average flux at different angles with respect to nadir.
Momenta dependence on magnetic latitude for (c)- downgoing and (d) - upgoing protons.}
\label{fig:506}
\end{center}
\end{figure}

\newpage

\begin{figure}[hp]
\begin{center}\mbox{
\epsfig{file=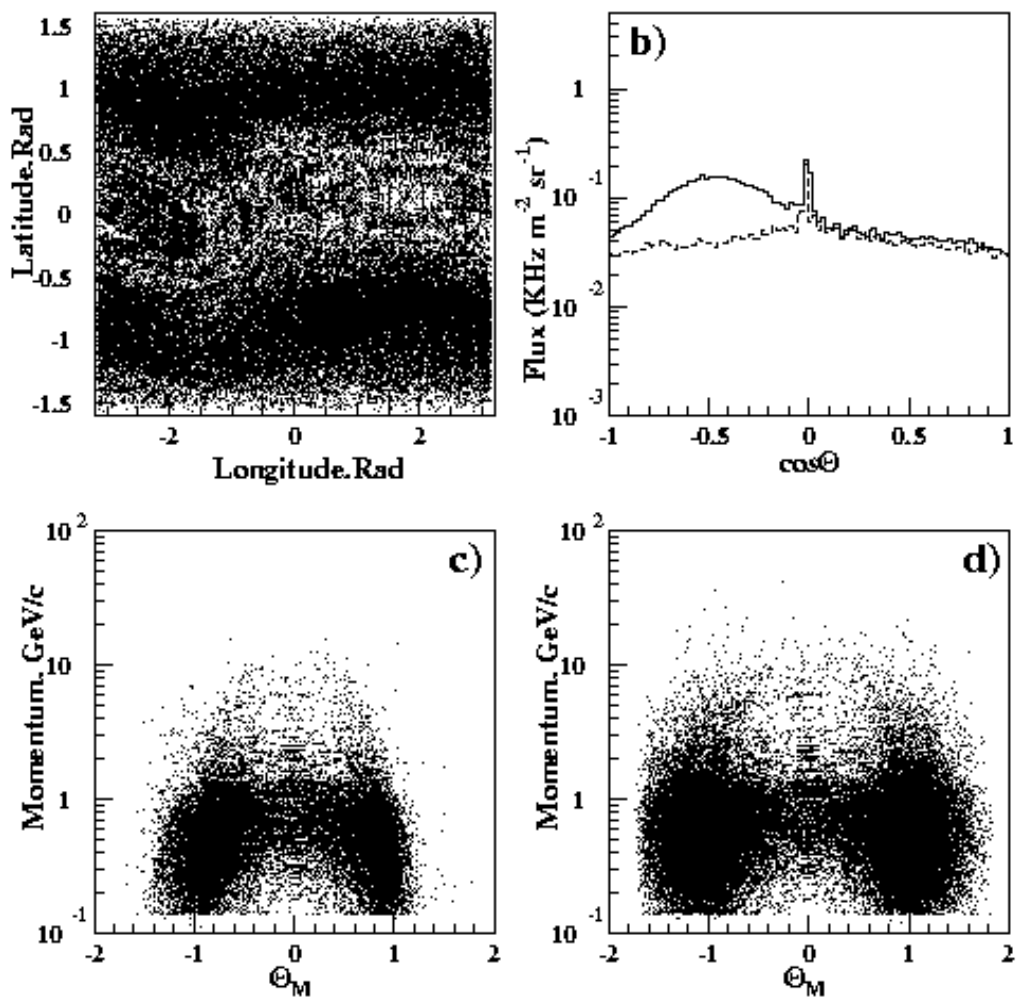,height=15 cm,width=15cm}}
\caption{%
Secondary protons detected at 400 km. 
(a) Geographic coordinates of the crossing points (b) Avergae flux at different angles with respect to 
nadir for pole ($|\Theta_M|>$0.7 rad, solid line) and equator ($|\Theta_M|<$0.7 rad., dashed line) 
regions, momentum dependence on  magnetic latitude for (c)- downgoing and (d) - upgoing protons.}
\label{fig:507}
\end{center}
\end{figure}

\newpage

\begin{figure}[hp]
\begin{center}\mbox{
\epsfig{file=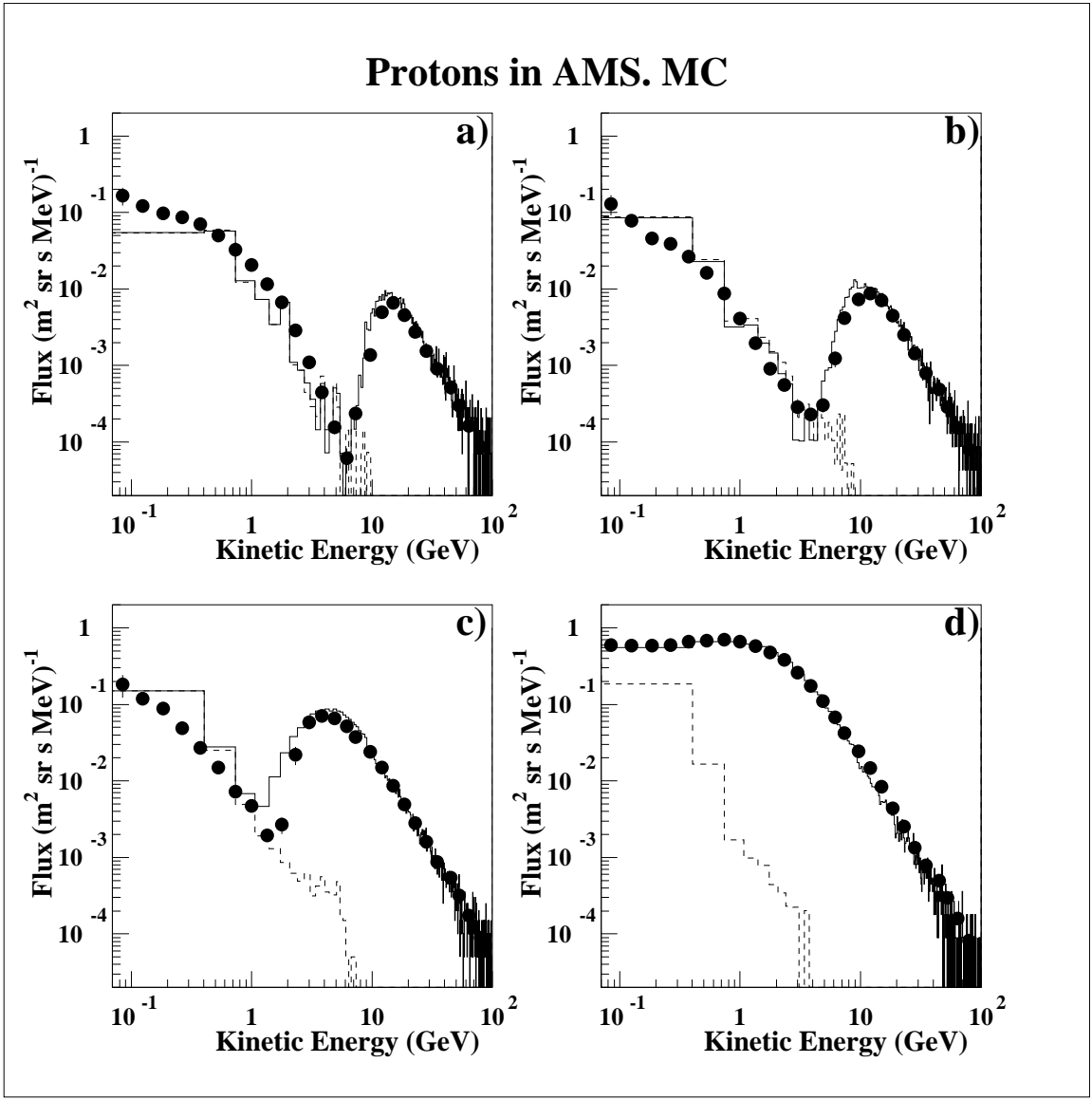,height=15 cm,width=15cm}}
\caption{%
Proton flux in AMS at different magnetic latitudes ($|\Theta_M|$). Solid line - flux from the top, 
dashed line - flux from the bottom.
a) $|\Theta_M|<$0.2 rad, b) 0.2$<|\Theta_M|<$0.5 rad, c) 0.5$<|\Theta_M|<$0.8 rad,
c) 0.8$<|\Theta_M|<$1.1 rad. The dots is the AMS measurement.}
\label{fig:508m}
\end{center}
\end{figure}

\newpage

\begin{figure}[hp]
\begin{center}\mbox{
\epsfig{file=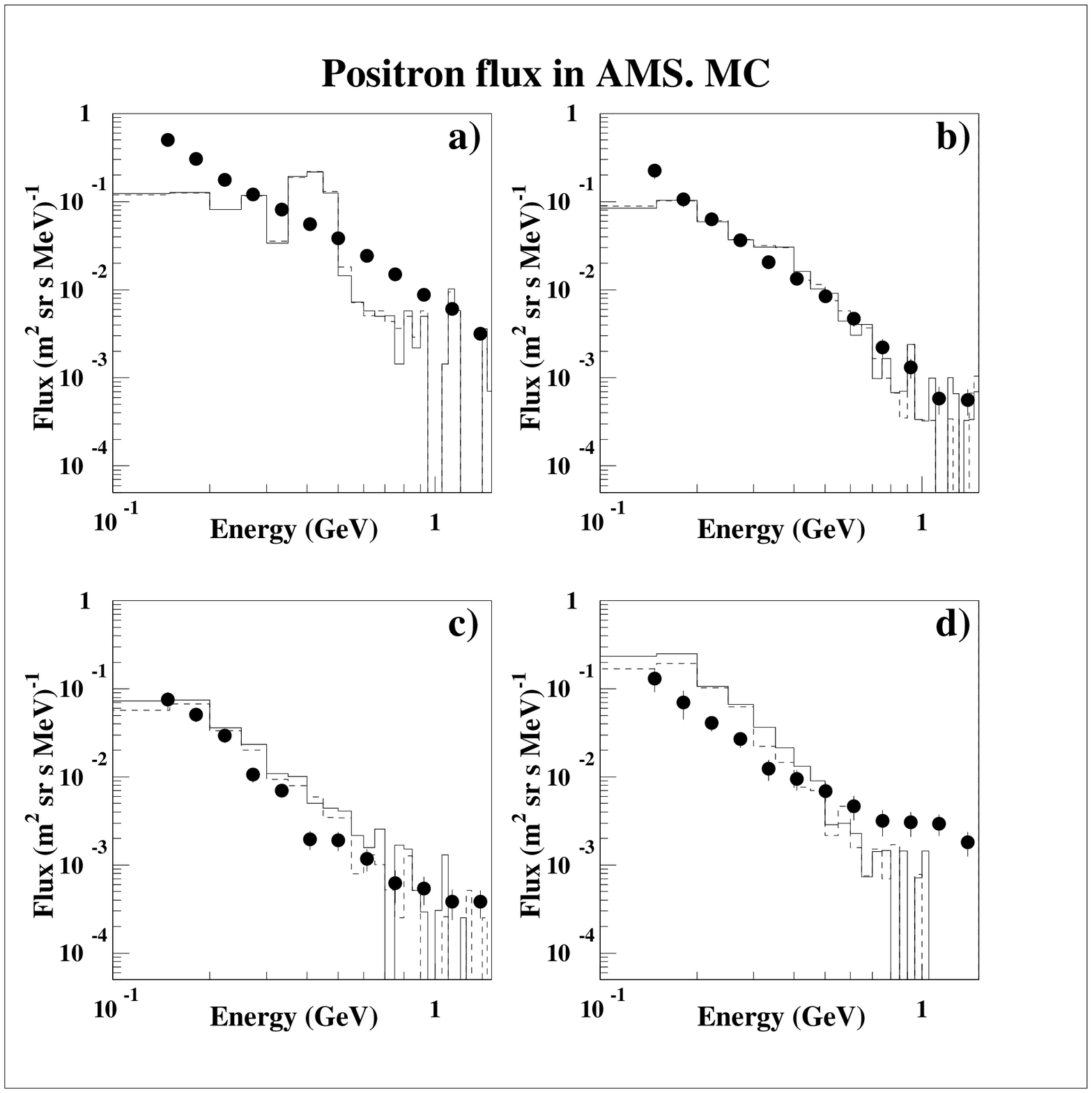,height=15 cm,width=15cm}}
\caption{%
Secondary positron flux in AMS at different magnetic latitudes ($|\Theta_M|$. Solid line - flux from the top, 
dashed line - flux from the bottom.
a) $|\Theta_M|<$0.3 rad, b) 0.3$<|\Theta_M|<$0.6 rad, c) 0.6$<|\Theta_M|<$0.8 rad,
c) 0.8$<|\Theta_M|<$1.1 rad. The dots is the AMS measurement. The enhancement at higher energies of the measured flux
at high magnetic latitude is due to primary cosmic positrons detected at the lower magnetic cutoff
region.}
\label{fig:510m}
\end{center}
\end{figure}

\newpage
\begin{figure}[hp]
\begin{center}\mbox{
\epsfig{file=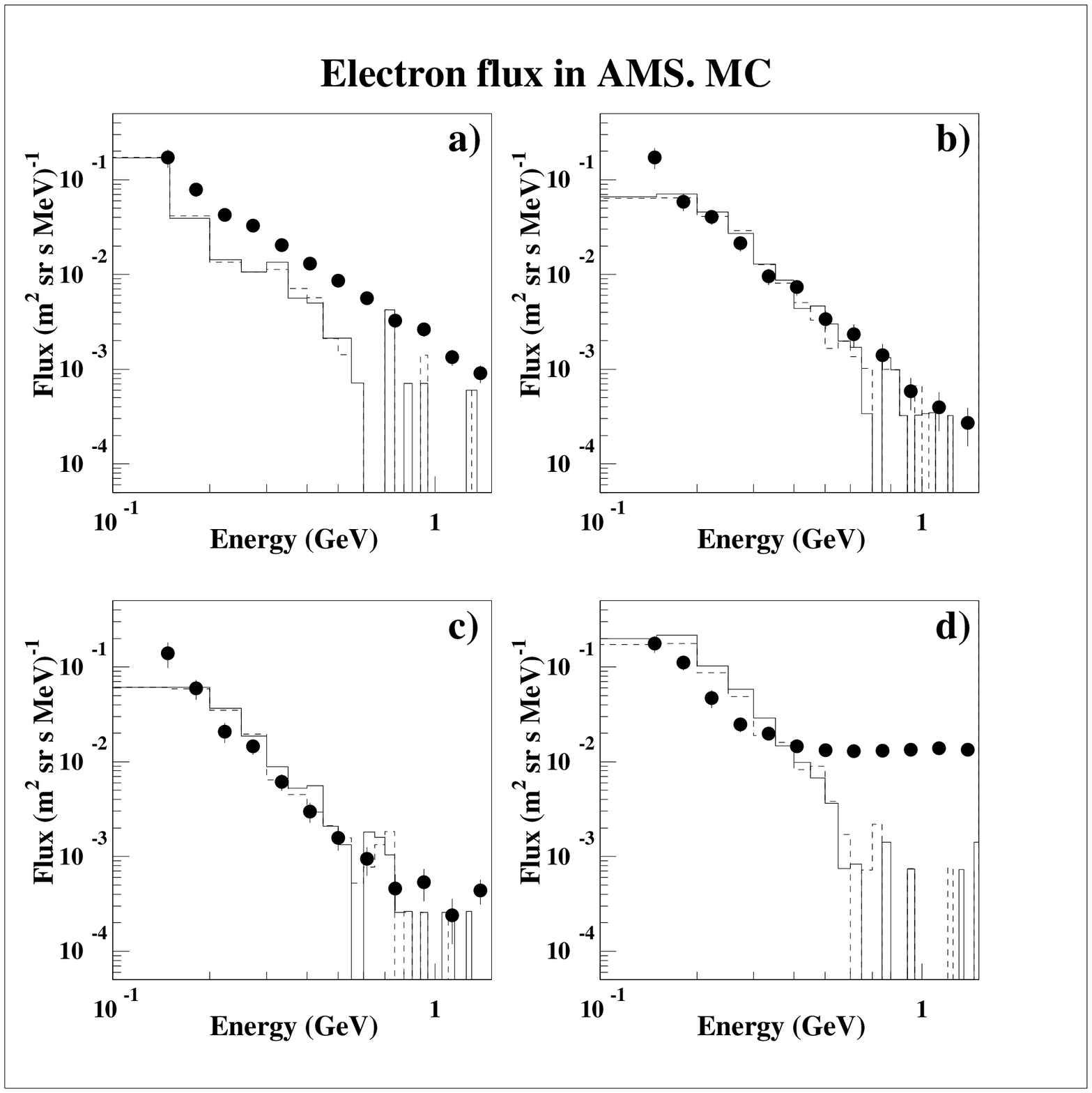,height=15 cm,width=15cm}}
\caption{%
Secondary electron flux in AMS at different magnetic latitudes ($|\Theta_M|$. Solid line - flux from the top, 
dashed line - flux from the bottom.
a) $|\Theta_M|<$0.3 rad, b) 0.3$<|\Theta_M|<$0.6 rad, c) 0.6$<|\Theta_M|<$0.8 rad,
d) 0.8$<|\Theta_M|<$1.1 rad. The dots is the AMS measurement. The enhancement at higher energies of the measured flux
at high magnetic latitude is due to primary cosmic electrons detected at the lower magnetic cutoff
region.}
\label{fig:511m}
\end{center}
\end{figure}

\newpage
\begin{figure}[ht]
\begin{center}
\mbox{\epsfig{file=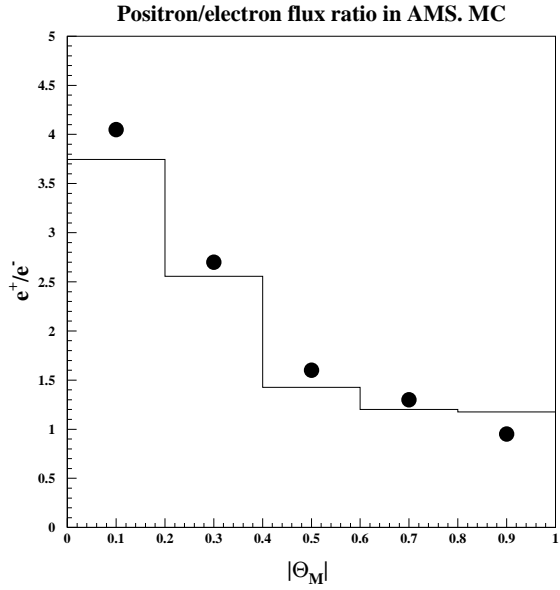,height=8.0 cm,width=8.0 cm}}
\caption{%
Calculated dependence on magnetic latitude of positron/electron flux ratio.
The dots is the AMS measurement.}
\label{fig:512m}
\end{center}
\end{figure}

\begin{figure}[hb]
\begin{center}\mbox{
\epsfig{file=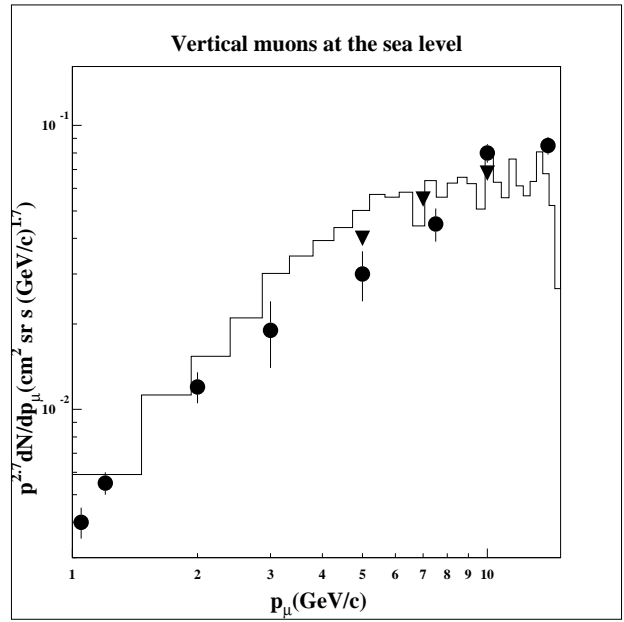,height=8 cm,width=8cm}}
\caption{%
Vertical muon flux at the sea level. Histrogram: this study. Circles: data from Ref.\cite{bib-mu1}, 
triangles: data from Ref.\cite{bib-mu2}.}  
\label{fig:muons}
\end{center}
\end{figure}

\newpage
  
  \item the energy spectra of under cutoff protons and their absolute flux.

\end{itemize}

Figures \ref{fig:510m} and \ref{fig:511m} show the spectra of secondary positrons and electrons
correspondingly. Although the positrons and electrons can not contribute to atmospheric neutrino
production they are (together with cosmic muons) the "spectators" of neutrino production in the 
decays of particles in the atmospheric showers and consequently the correct prediction of the 
fluxes of these particles is a proof of validity of the whole simulation procedure including the 
neutrino flux predictions. The simulated positron and electron spectra are similar to those 
obtained in the AMS experiment \cite{bib-AMS2} for the secondary leptons in terms of equality 
of fluxes of upgoing and downgoing particles and energy dependence of the flux at different 
magnetic latitudes.

 The difference in the measured and simulated secondary leptonic (and to some 
extent secondary proton) fluxes at low magnetic latitudes
(Figs.\ref{fig:510m}a, \ref{fig:511m}a and \ref{fig:508m}a) is due to insufficient statistics of 
the present Monte Carlo study. As mentioned above, the measured rate in this region is largerly 
connected with the relatively rare long traveling particles multiply crossing the AMS orbit plane.
Actually, the peak in the spectrum of figure \ref{fig:510m}a is due to just 3 spiraling positrons 
each crossing the AMS detector plane about 500 times in both upward and downward direction.   
Nevertheless the measured relative rate of secondary leptons \cite{bib-AMS2,bib-AMS4}
and their dependence on the magnetic latitude is well reproduced by the simulation 
(Fig.\ref{fig:512m}). 
It is worth to emphasize that the impact of long traveling secondaries on the production of 
atmospheric neutrinos is negligible.  

Apart from the data connected with those measured by AMS the present simulation includes a prediction of the
fluxes of muons at the sea level. The histogram in Fig.\ref{fig:muons} shows a result obtained
in this study for the vertical muon flux. The result is in good agreement with the measured fluxes 
reported in \cite{bib-mu1,bib-mu2}.

\section{\bf Atmospheric neutrino fluxes}
\begin{figure}[hp]
\begin{center}\mbox{
\epsfig{file=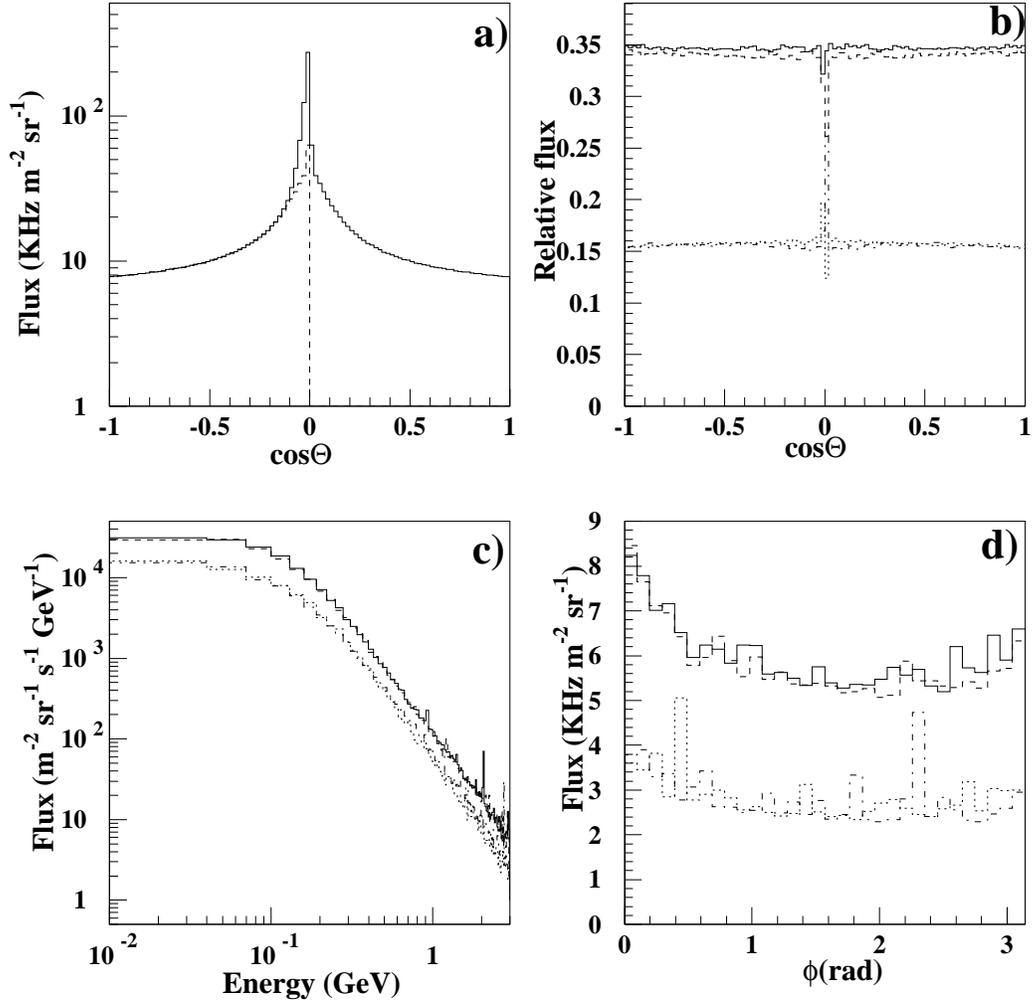,height=15 cm,width=15cm}}
\caption{%
Neutrino flux averaged over the Earth's surface. The angle of incidence ($\Theta$) is measured
from nadir (cos$\Theta$=1 for particles going vertically down). The dashed line in a) is a mirror
image of the flux from the top. b) Relative flux of different kinds of neutrino. c) Energy spectra.
d) Azimuth angle measured from direction toward east.
Solid line - $\tilde{\nu}_{\mu}$, dashed line - $\nu_{\mu}$, dotted line - $\tilde{\nu}_{e}$, 
dashed-dotted line - $\nu_{e}$.} 
\label{fig:514}
\end{center}
\end{figure}

As for the fluxes of charged particles discussed in the previous section all results that
follow are obtained directly without any renormalization. 
The dependence of the neutrino flux averaged over the entire Earth's surface on the neutrino
direction with respect to nadir is presented in Fig.\ref{fig:514}a. The overall equality
of neutrino fluxes to and from the Earth is clearly seen. The dashed line in the figure is a 
mirror image of the flux from the top. A small increase of the flux of almost horizontal neutrinos 
coming from inside the Earth is due to neutrinos produced by the cosmic particle's showers in the Earth.
The relative flux of different kinds of atmospheric neutrinos
($\nu_{\mu},\tilde{\nu}_{\mu},\nu_{e},\tilde{\nu}_{e}$) is constant for all incidence angles 
(Fig.\ref{fig:514}b.). The averaged over all directions energy spectra of neutrino of different kind 
are presented in  Fig.\ref{fig:514}c. 
The dependence of the flux on the
azimuth (with respect to east direction) angle (Fig.\ref{fig:514}d) reflects some east-west asymmetry
of average neutrino fluxes. 

\newpage

\newpage

\begin{figure}[hb]
\begin{center}\mbox{
\epsfig{file=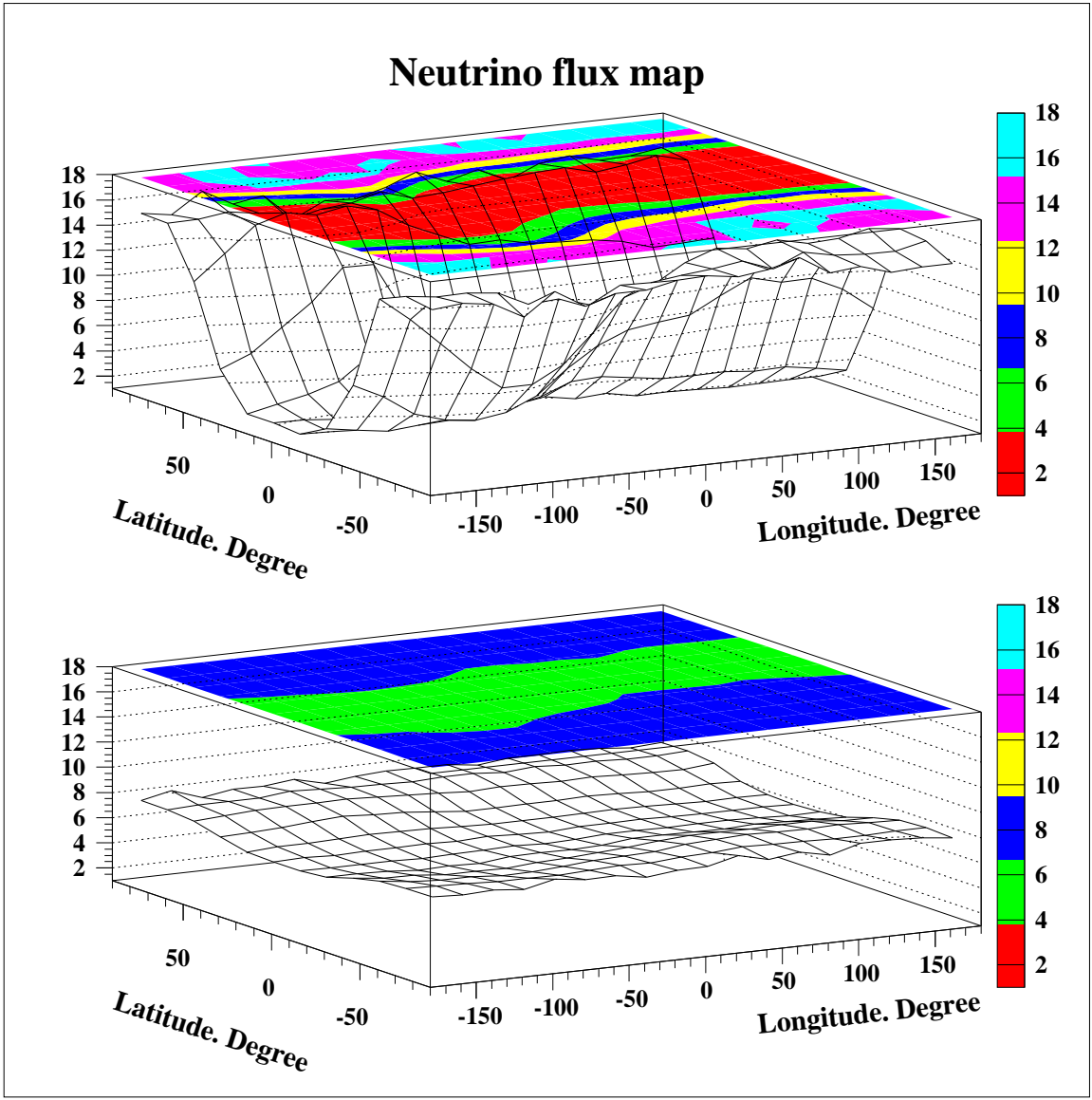,height=15 cm,width=15cm}}
\caption{%
The geographical map of the average atmospheric neutrino flux in KHz/$\rm m^2sr$. 
Top: Downward doing. Bottom:Upward going.}
\label{fig:521}
\end{center}
\end{figure}
\newpage

The actual picture of neutrino flux distribution is very different from the average one.
Fig.\ref{fig:521} shows the geographic maps of neutrino fluxes measured in KHz/$\rm m^2sr$ for downgoing
(top) and upgoing (bottom) atmospheric neutrinos. The downward flux is very nonuniform and, not 
surprisingly, strongly 
correlated with the flux of cosmic protons (see fig.\ref{fig:501}). The downward neutrino flux changes 
from more than 15 KHz/$\rm m^2sr$ at the magnetic poles down to less than 3 KHz/$\rm m^2sr$ at the 
magnetic equator
whereas the upward flux is almost uniform at the level of 6-7 KHz/$\rm m^2sr$ over the entire Earth's 
surface.  For different regions of the Earth not only the total flux of neutrinos is different but 
also the kinematical parameters of the neutrinos. 

To allow the comparison with previous calculations of Ref.\cite{bib-LipGeo} we compute the neutrino 
event rate in three equal area regions according to the geomagnetic latitude, although in view of 
neutrino flux rapidly changing with magnetic latitude (see Fig.\ref{fig:521}) such division appears
to be rather crude. 

\begin{figure}[hb]
\begin{center}\mbox{
\epsfig{file=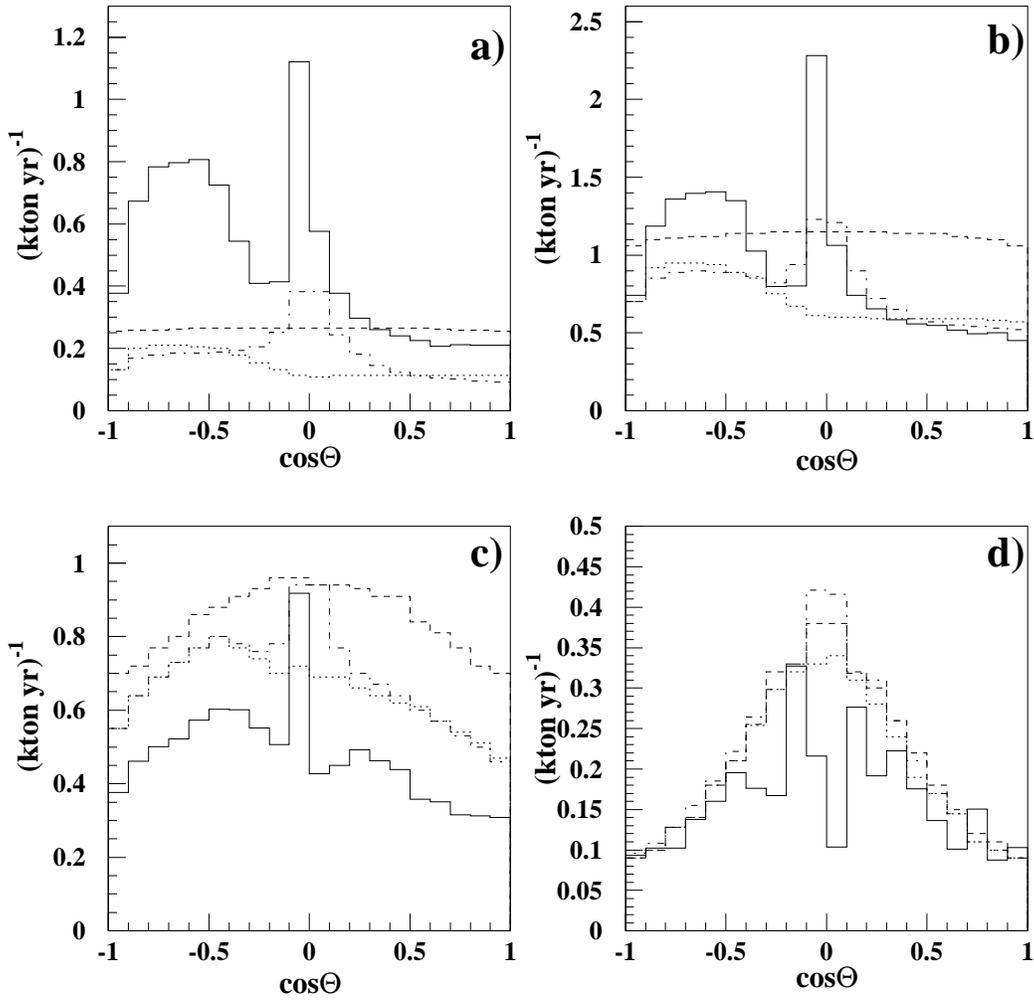,height=15 cm,width=16.3cm}}
\caption{%
Average nadir angle distribution for e-like events for detectors located in positions on the Earth  
with magnetic latitude sin($\rm \Theta_M$)= [-0.2,0.2]. a) $\rm E_{\nu}$=0.1-0.31 GeV, 
b) $\rm E_{\nu}$=0.31-1.0 GeV, c) $\rm E_{\nu}$=1.0-3.1 GeV, d) $\rm E_{\nu}$=3.1-10.0 GeV.
From Ref.\cite{bib-LipGeo} : dashed line - 1D calculation without geomagnetic effects; 
dotted line - 1D calculation; dash-dotted line - 3D calculation. Solid line -this work.}
\label{fig:522}
\end{center}
\end{figure}

\begin{figure}[hb]
\begin{center}\mbox{
\epsfig{file=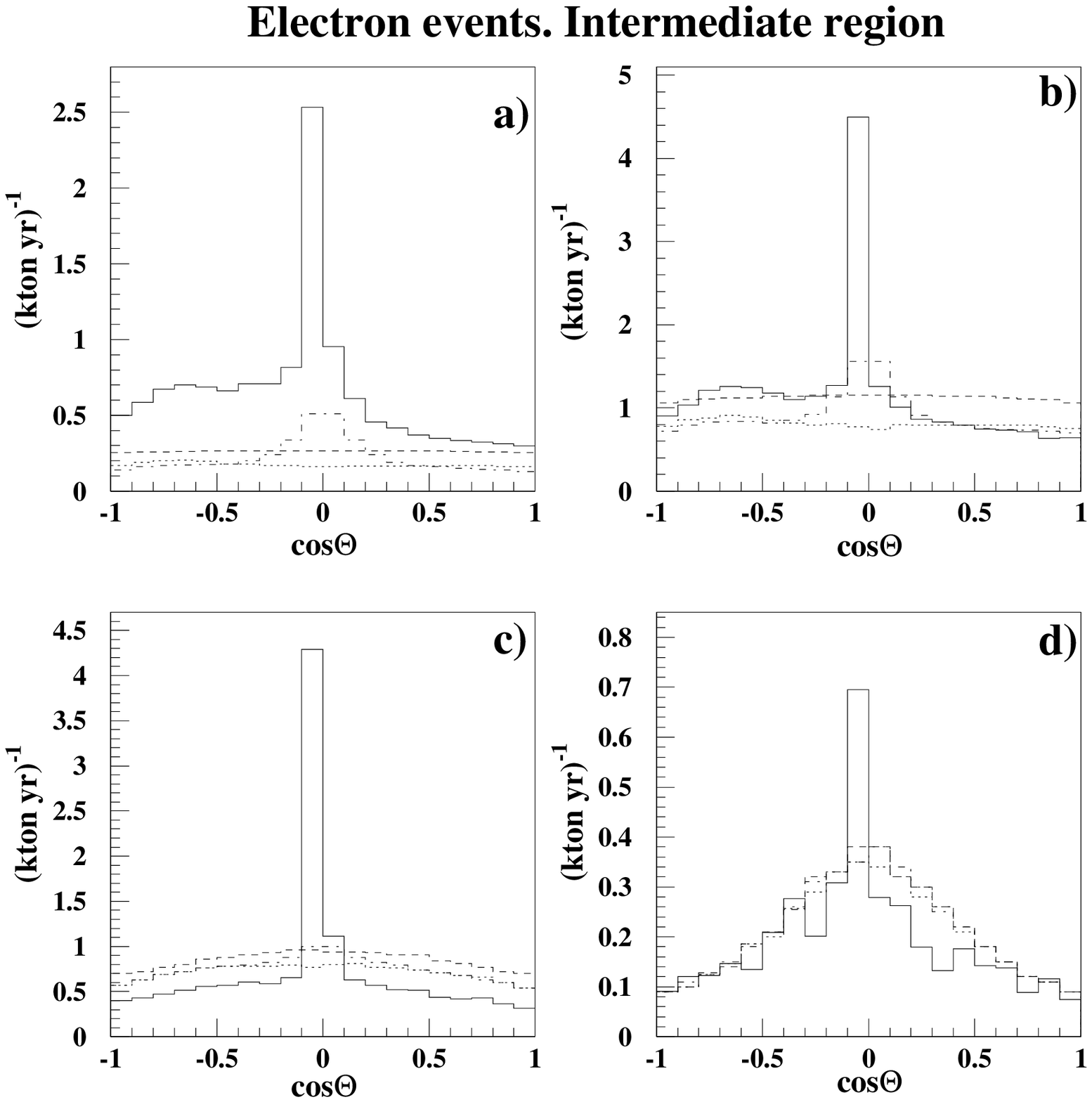,height=15 cm,width=16.3cm}}
\caption{%
Average nadir angle distribution for e-like events for detectors located in positions on the Earth  
with magnetic latitude sin($\rm \Theta_M$)= [0.2,0.6]. a) $\rm E_{\nu}$=0.1-0.31 GeV, 
b) $\rm E_{\nu}$=0.31-1.0 GeV, c) $\rm E_{\nu}$=1.0-3.1 GeV, d) $\rm E_{\nu}$=3.1-10.0 GeV.
From Ref.\cite{bib-LipGeo} : dashed line - 1D calculation without geomagnetic effects; 
dotted line - 1D calculation; dash-dotted line - 3D calculation. Solid line -this work.}
\label{fig:523}
\end{center}
\end{figure}

\begin{figure}[hb]
\begin{center}\mbox{
\epsfig{file=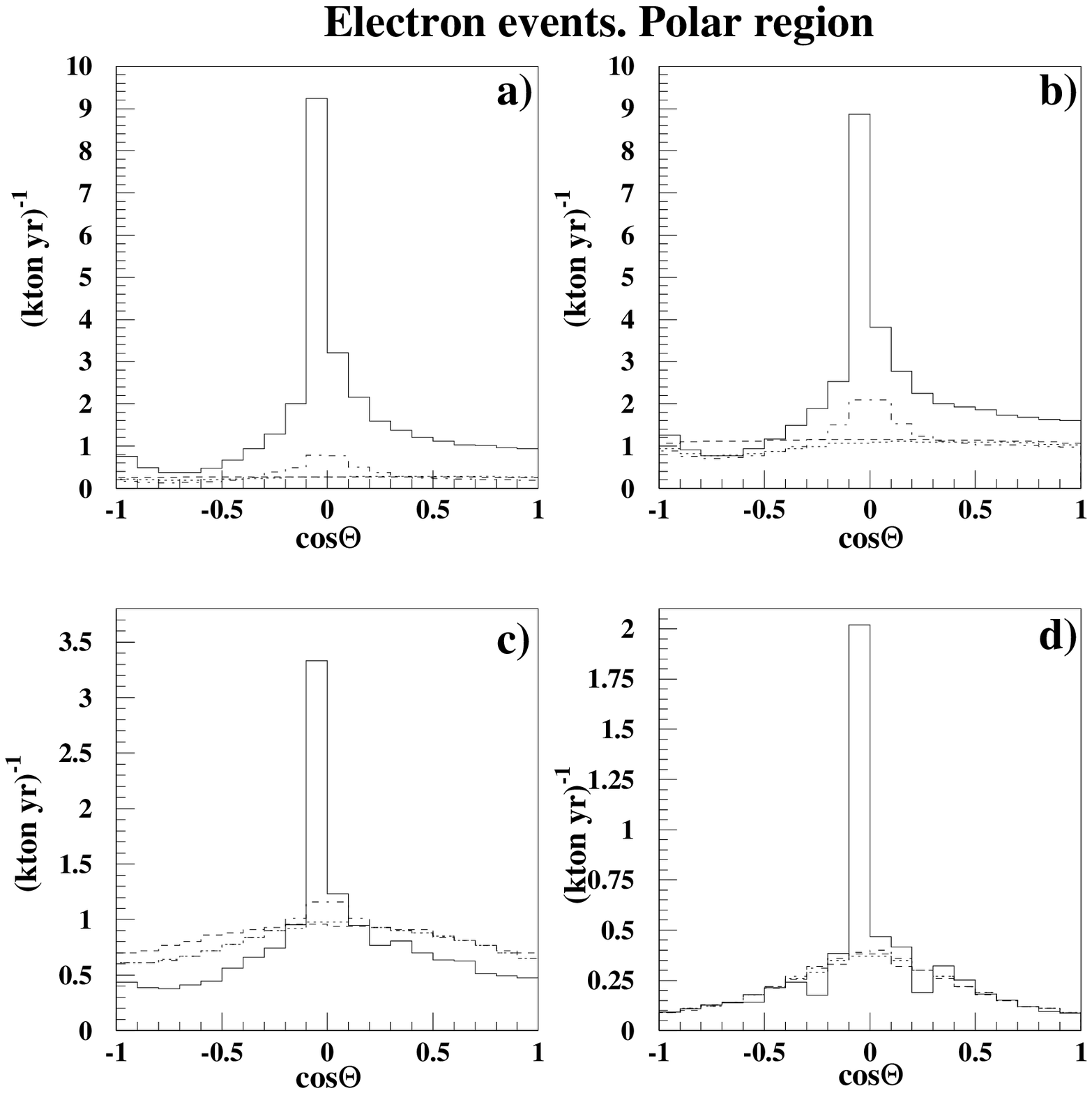,height=15 cm,width=16.3cm}}
\caption{%
Average nadir angle distribution for e-like events for detectors located in positions on the Earth  
with magnetic latitude sin($\rm \Theta_M$)= [0.6,1.0]. a) $\rm E_{\nu}$=0.1-0.31 GeV, 
b) $\rm E_{\nu}$=0.31-1.0 GeV, c) $\rm E_{\nu}$=1.0-3.1 GeV, d) $\rm E_{\nu}$=3.1-10.0 GeV.
From Ref.\cite{bib-LipGeo} : dashed line - 1D calculation without geomagnetic effects; 
dotted line - 1D calculation; dash-dotted line - 3D calculation. Solid line -this work.}
\label{fig:524}
\end{center}
\end{figure}

\begin{figure}[hb]
\begin{center}\mbox{
\epsfig{file=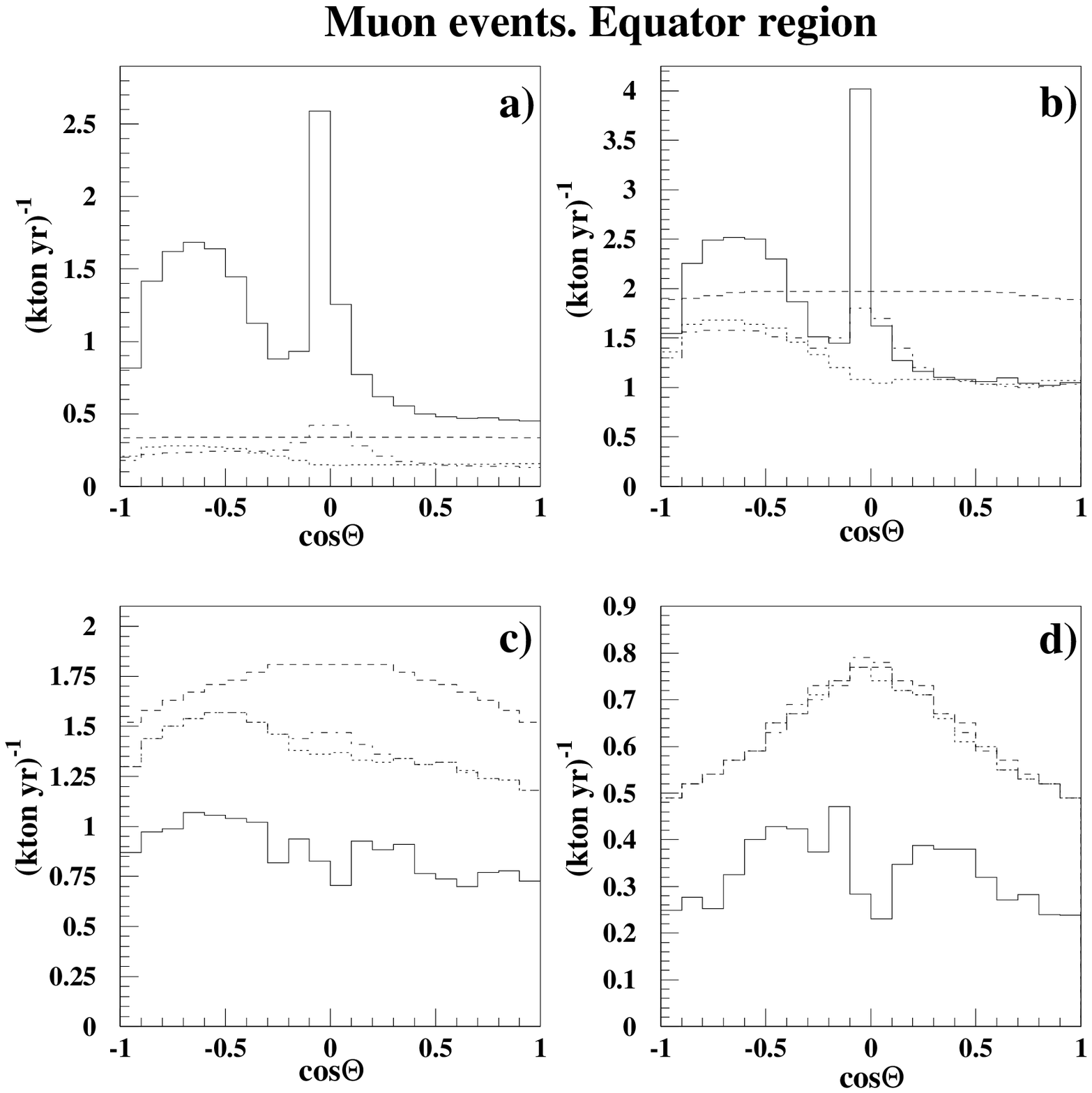,height=15 cm,width=16.3cm}}
\caption{%
Average nadir angle distribution for $\mu$-like events for detectors located in positions on the Earth  
with magnetic latitude sin($\rm \Theta_M$)= [-0.2,0.2]. a) $\rm E_{\nu}$=0.1-0.31 GeV, 
b) $\rm E_{\nu}$=0.31-1.0 GeV, c) $\rm E_{\nu}$=1.0-3.1 GeV, d) $\rm E_{\nu}$=3.1-10.0 GeV.
From Ref.\cite{bib-LipGeo} : dashed line - 1D calculation without geomagnetic effects; 
dotted line - 1D calculation; dash-dotted line - 3D calculation. Solid line -this work.}
\label{fig:525}
\end{center}
\end{figure}

\begin{figure}[hb]
\begin{center}\mbox{
\epsfig{file=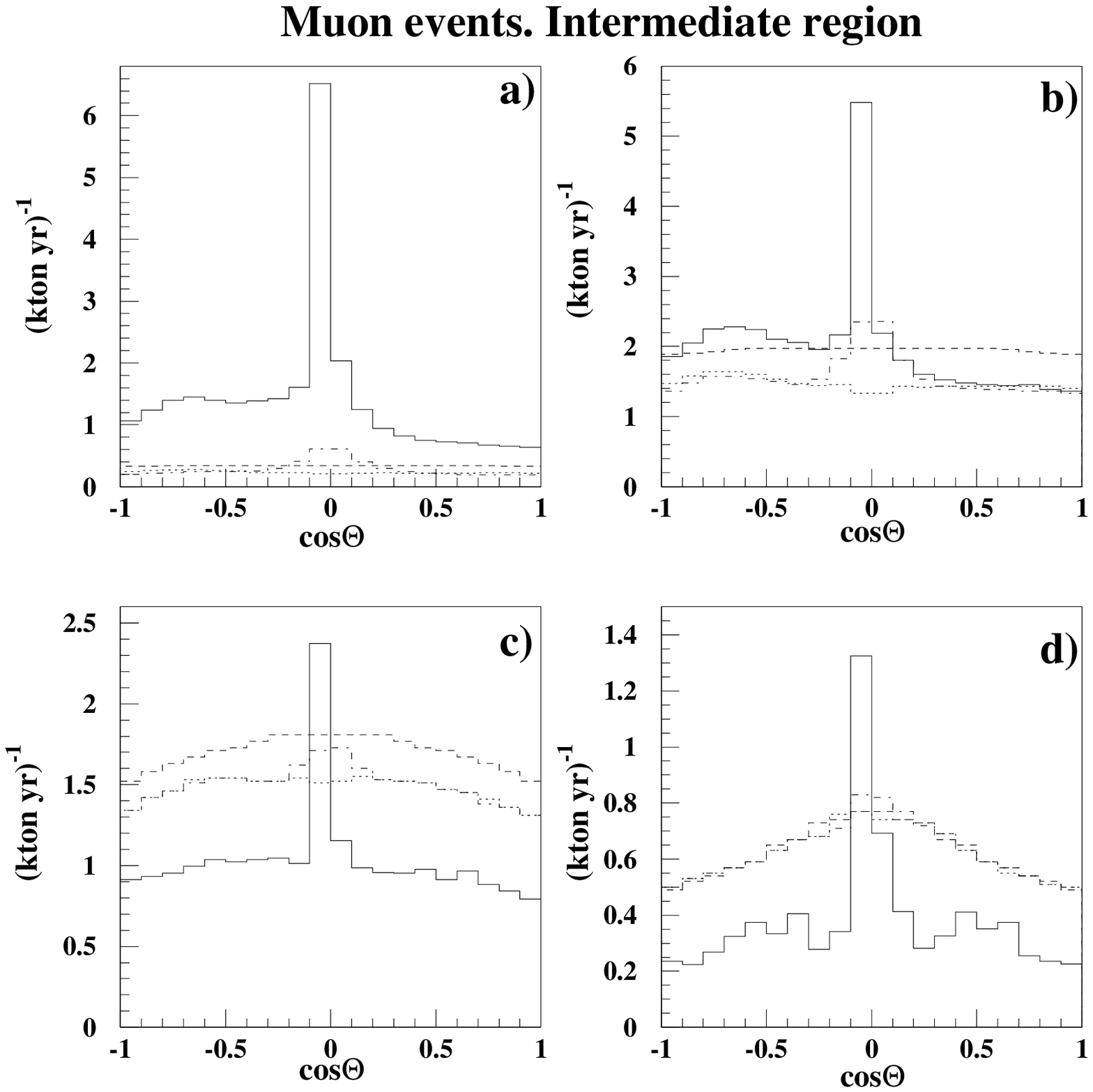,height=15 cm,width=16.3cm}}
\caption{%
Average nadir angle distribution for $\mu$-like events for detectors located in positions on the Earth  
with magnetic latitude sin($\rm \Theta_M$)= [0.2,0.6]. a) $\rm E_{\nu}$=0.1-0.31 GeV, 
b) $\rm E_{\nu}$=0.31-1.0 GeV, c) $\rm E_{\nu}$=1.0-3.1 GeV, d) $\rm E_{\nu}$=3.1-10.0 GeV.
From Ref.\cite{bib-LipGeo} : dashed line - 1D calculation without geomagnetic effects; 
dotted line - 1D calculation; dash-dotted line - 3D calculation. Solid line -this work.}
\label{fig:526}
\end{center}
\end{figure}

\begin{figure}[hb]
\begin{center}\mbox{
\epsfig{file=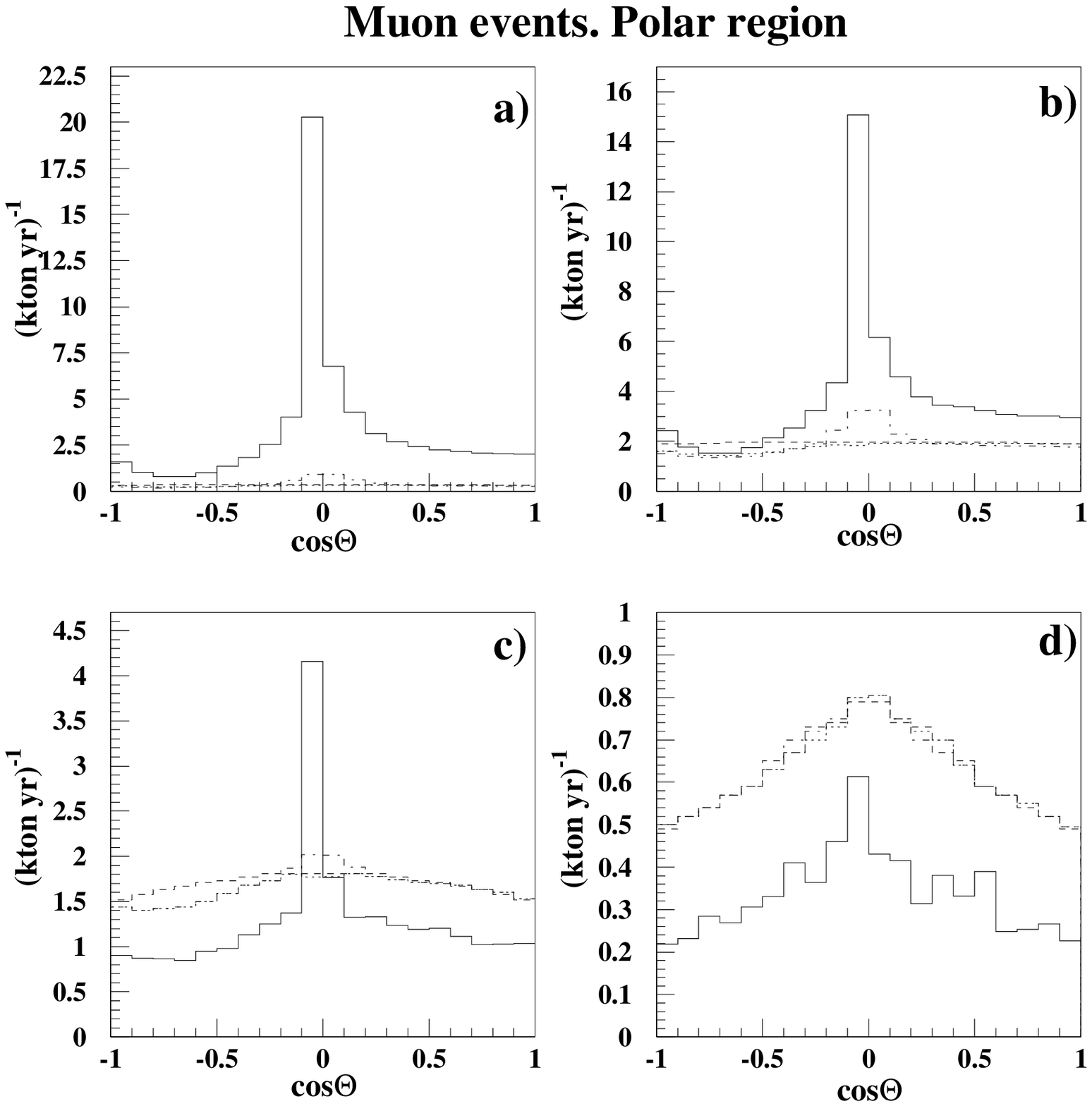,height=15 cm,width=16.3cm}}
\caption{%
Average nadir angle distribution for $\mu$-like events for detectors located in positions on the Earth  
with magnetic latitude sin($\rm \Theta_M$)= [0.6,1.0]. a) $\rm E_{\nu}$=0.1-0.31 GeV, 
b) $\rm E_{\nu}$=0.31-1.0 GeV, c) $\rm E_{\nu}$=1.0-3.1 GeV, d) $\rm E_{\nu}$=3.1-10.0 GeV.
From Ref.\cite{bib-LipGeo} : dashed line - 1D calculation without geomagnetic effects; 
dotted line - 1D calculation; dash-dotted line - 3D calculation. Solid line -this work.}
\label{fig:527}
\end{center}
\end{figure}

As in \cite{bib-LipGeo}
to compute the event rates the $\nu$ fluxes have been convoluted with the neutrino cross section
\cite{bib-LipConv},\cite{bib-Conv1},\cite{bib-Conv2}. The angle in the distributions 
(Figs.\ref{fig:522},\ref{fig:523},\ref{fig:524},\ref{fig:525},\ref{fig:526},\ref{fig:527})
is the neutrino one with no experimental smearing or inefficiency included. Apparently, 
the results of the present study differ from previous calculations both in the energy spectra and 
the angular distributions of atmospheric neutrinos. The flux of sub-GeV neutrino predicted by this 
calculation depends much stronger on the geographical region following the intensity of the primary
cosmic particle flux (Fig.\ref{fig:501}). The down/up asymmetry of the flux is much stronger.       
There is an important increase in the horizontal flux emerging from the Earth. This flux is connected with
neutrinos produced under the Earth's surface at a depth of several meters. A part of these 
neutrinos can be detected in the underground detectors. The actual amount of "underground" neutrinos
detected in a given detector depends on the detector position depth.

\section{\bf Atmospheric neutrino fluxes at major experimental sites}

\begin{figure}[hp]
\begin{center}\mbox{
\epsfig{file=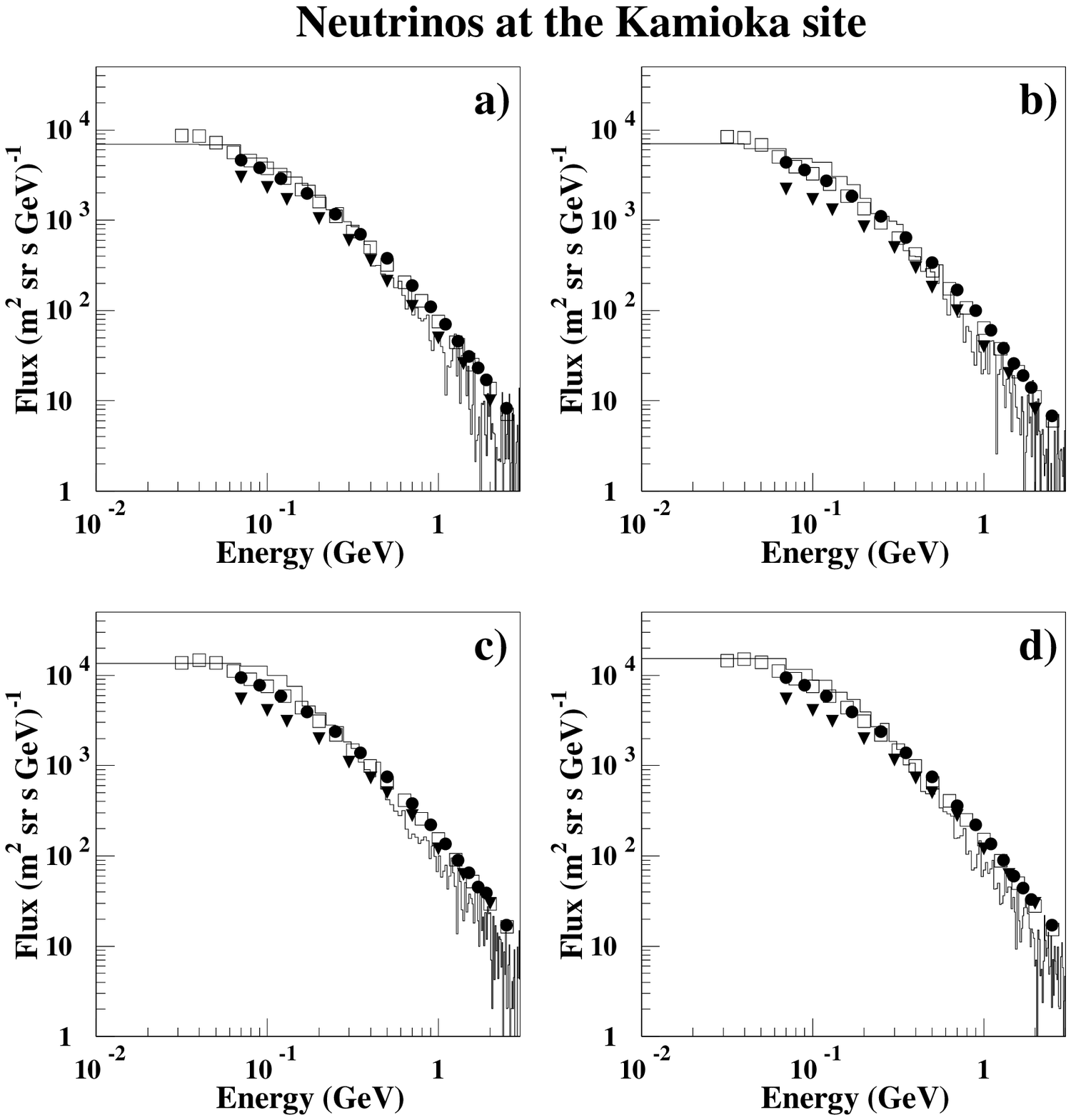,height=15 cm,width=16.3cm}}
\caption{%
Differential energy spectra of atmospheric neutrino (averaged over all directions) 
at the Kamioka site. 
a) $\nu_{e}$, b) $\tilde{\nu}_{e}$, c) $\nu_{\mu}$, d) $\tilde{\nu}_{\mu}$. 
Squares from Ref. \cite{bib-Hon},
dots from Ref. \cite{bib-Barr}, triangles from Ref. \cite{bib-Bug}. Histograms - this work.}
\label{fig:515a}
\end{center}
\end{figure}

\begin{figure}[hp]
\begin{center}\mbox{
\epsfig{file=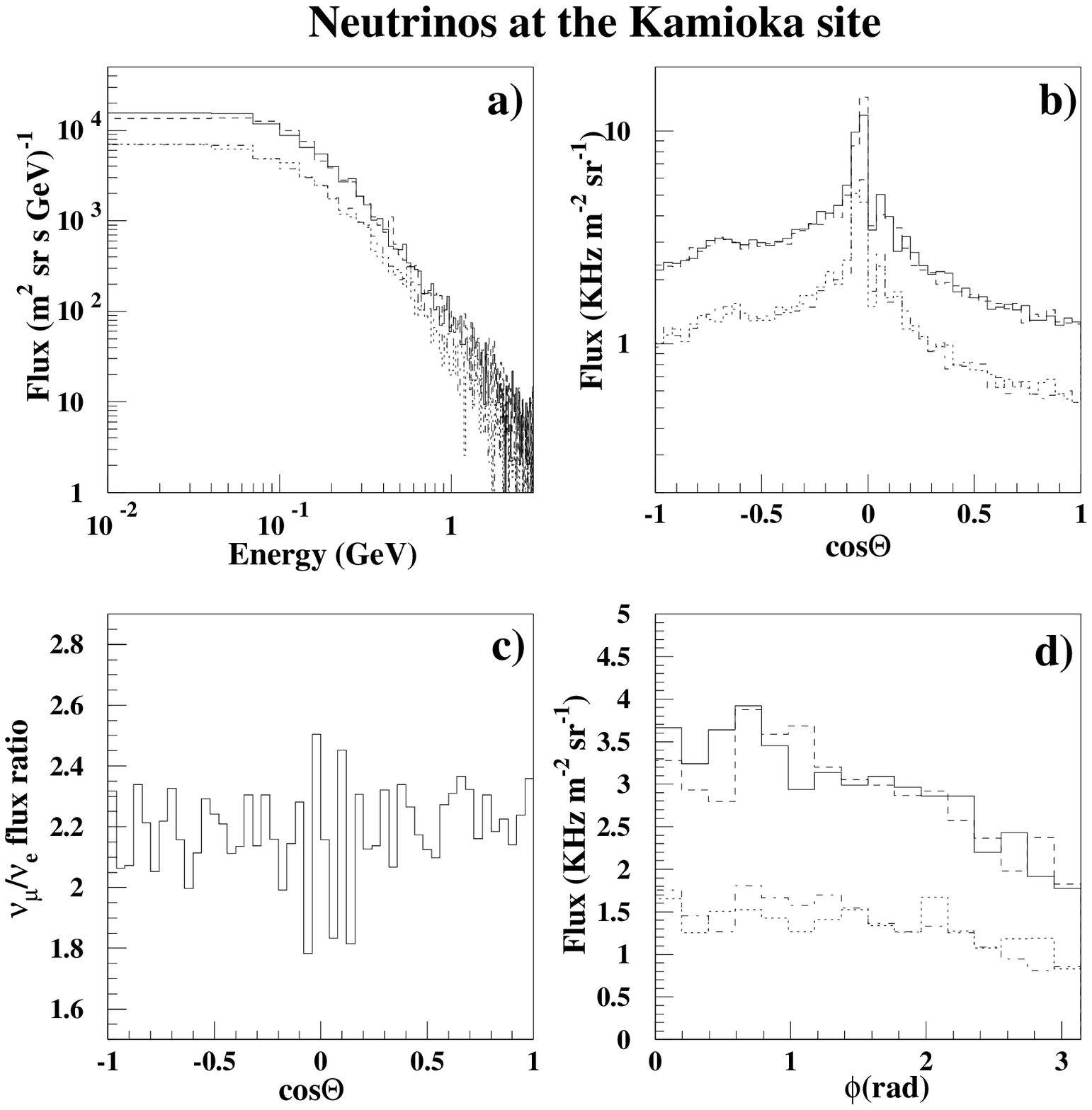,height=15 cm,width=16.3cm}}
\caption{%
Neutrino flux at the Kamioka site. 
a) Energy spectra. b) Nadir angle distribution. c) Flux ratio of $\mu/e$ neutrinos.
d) Azimuth angle distribution.
In a),b) and d) solid line - $\tilde{\nu}_{\mu}$, dashed line - $\nu_{\mu}$, 
dotted line - $\tilde{\nu}_{e}$, 
dashed-dotted line - $\nu_{e}$.} 
\label{fig:515}
\end{center}
\end{figure}

For the relatively small area corresponding to different experimental sites 
the statistics of the present study is high enough to make reliable conclusions for neutrinos 
with energies below $\sim$1.5 GeV.
The parameters describing the neutrino flux at the Kamioka site are calculated for an
area within the geographical coordinates $35^o\pm12^o~ N,~142^o\pm12^o~ E$ for an underground detector 
at a depth of 520 m (2700 mwe). 
The averaged over all directions differential energy spectra of different neutrinos and antineutrinos 
are given in Fig.\ref{fig:515a} together with the results from other authors. The present study gives 
lower neutrino rates at higher energies. 
The fluxes of all kinds of neutrino are strongly down/up asymmetric with a considerable contribution
of "underground" neutrinos in the horizontal direction (Fig.\ref{fig:515}b).
The ratio of muon($\nu_{\mu}+\tilde{\nu}_{\mu}$) to electron($\nu_{e}+\tilde{\nu}_{e}$) fluxes 
(Fig.\ref{fig:515}c) suggests a tendency to decrease for horizontal directions.

\newpage

\newpage

\newpage

\newpage

\newpage

\newpage

\newpage

\newpage
\begin{figure}[hp]
\begin{center}\mbox{
\epsfig{file=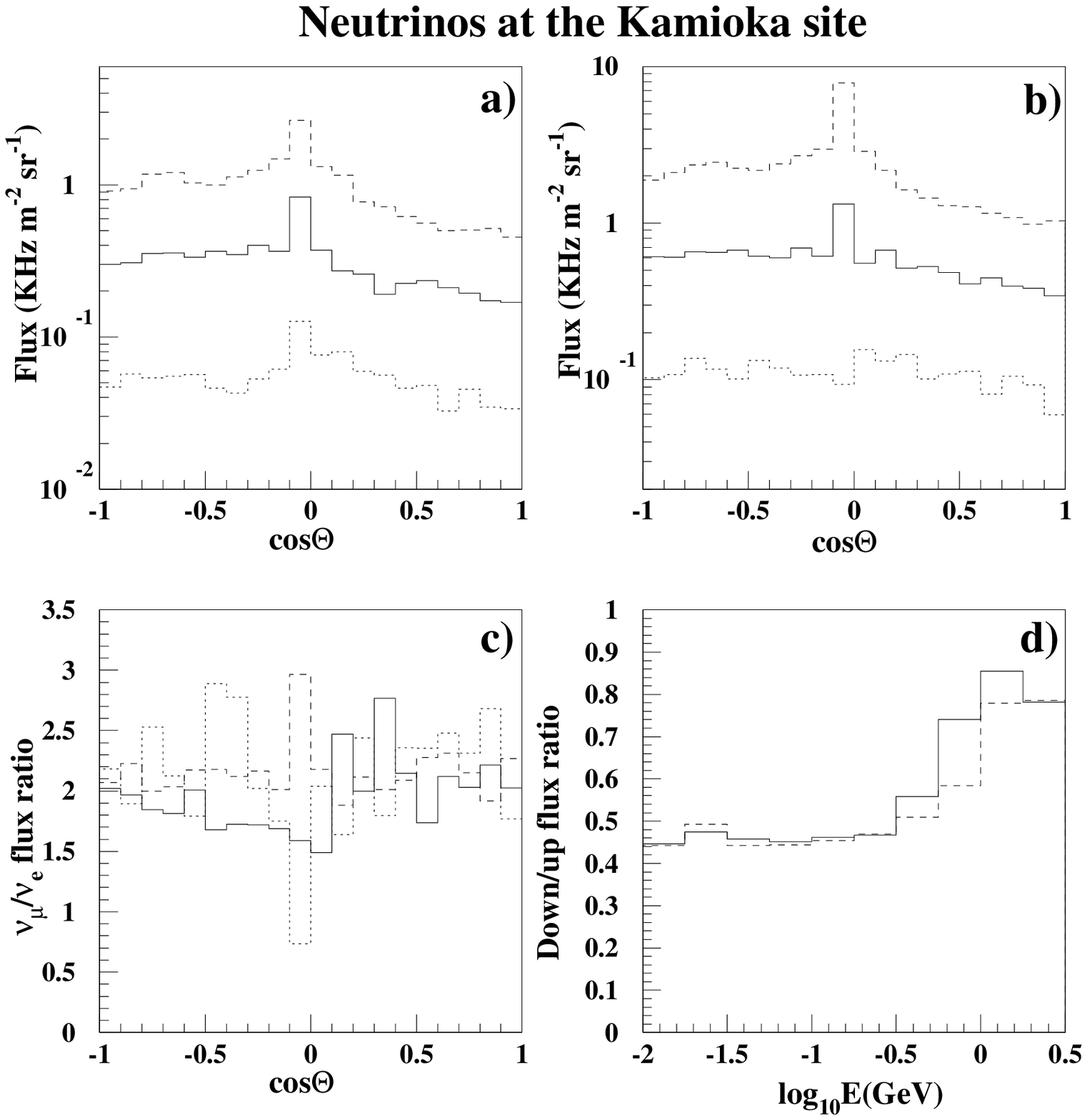,height=15 cm,width=16.3cm}}
\caption{%
Neutrino flux at the Kamioka site. 
Nadir angle distribution for a) electron ($\nu_{e},\tilde{\nu}_{e}$) and b)
muon ($\nu_{\mu},\tilde{\nu}_{\mu}$) neutrinos of different energy.
d) Energy dependence of the ratio of 
the muon/electron fluxes. In a),b) and c) - dashed line 0.1$<E<$0.31 GeV, solid line 0.31$<E<$1.0 GeV,
dotted line $E>$1.0 GeV.
d) Energy dependence of the flux ratio of downgoing to upgoing muon (solid line) and
electron (dashed line) neutrinos.}
\label{fig:516}
\end{center}
\end{figure}

\newpage

\begin{figure}[hp]
\begin{center}\mbox{
\epsfig{file=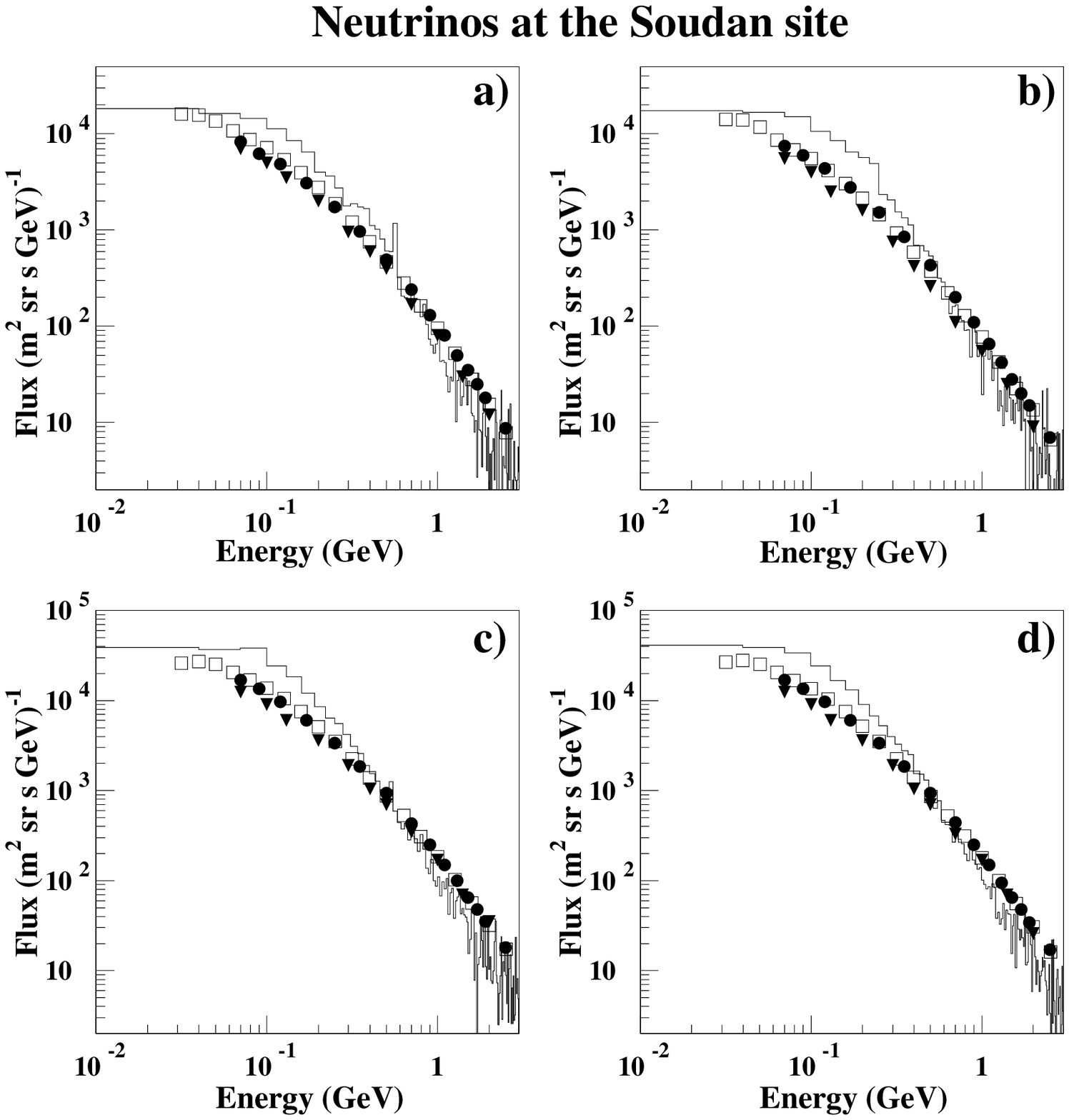,height=15 cm,width=16.3cm}}
\caption{%
Differential energy spectra of atmospheric neutrino (averaged over all directions)
 at the Saudan site. 
a) $\nu_{e}$, b) $\tilde{\nu}_{e}$, c) $\nu_{\mu}$, d) $\tilde{\nu}_{\mu}$.
Squares from Ref.\cite{bib-Hon},
dots from Ref.\cite{bib-Barr}, triangles from Ref.\cite{bib-Bug}. Histograms - this work.}
\label{fig:517a}
\end{center}
\end{figure}

\newpage
\begin{figure}[hp]
\begin{center}\mbox{
\epsfig{file=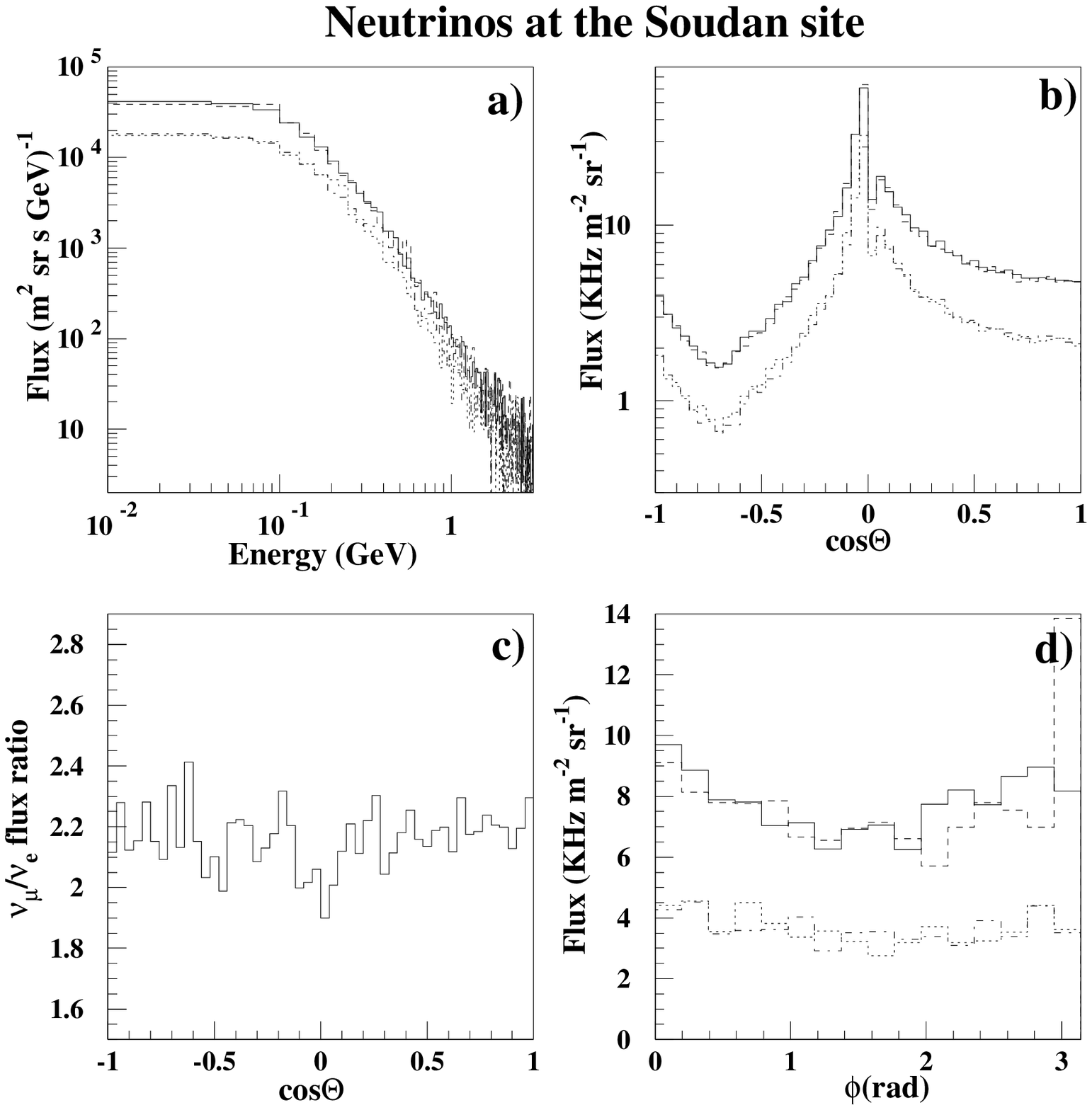,height=15 cm,width=16.3cm}}
\caption{%
Neutrino flux at the Soudan site. 
a) Energy spectra. b) Nadir angle distribution. c)Flux ratio of $\mu/e$ neutrinos.
d) Azimuth angle distribution.
In a),b) and d) solid line - $\tilde{\nu}_{\mu}$, dashed line - $\nu_{\mu}$,
dotted line - $\tilde{\nu}_{e}$, 
dashed-dotted line - $\nu_{e}$.} 
\label{fig:517}
\end{center}
\end{figure}

\newpage
\begin{figure}[hp]
\begin{center}\mbox{
\epsfig{file=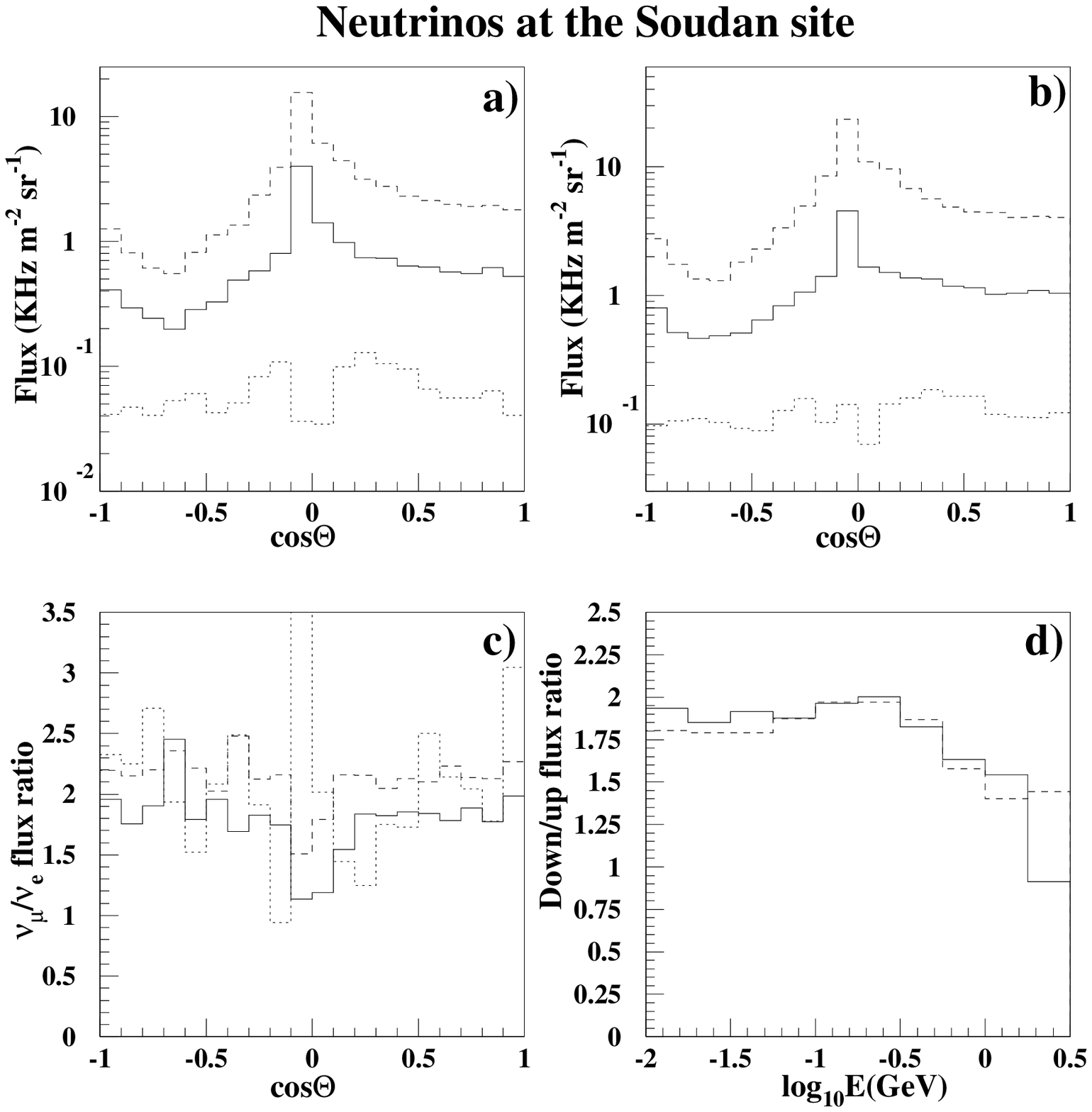,height=15 cm,width=16.3cm}}
\caption{%
Neutrino flux at the Soudan site. 
Nadir angle distribution for a) electron ($\nu_{e},\tilde{\nu}_{e}$) and b)
muon ($\nu_{\mu},\tilde{\nu}_{\mu}$) neutrinos of different energy.
d) Energy dependence of the ratio of 
the muon/electron fluxes. In a),b) and c) - dashed line 0.1$<E<$0.31 GeV, solid line 0.31$<E<$1.0 GeV,
dotted line $E>$1.0 GeV.
d) Energy dependence of the flux ratio of downgoing to upgoing muon (solid line) and
electron (dashed line) neutrinos.}
\label{fig:518}
\end{center}
\end{figure}

\newpage

\begin{figure}[hp]
\begin{center}\mbox{
\epsfig{file=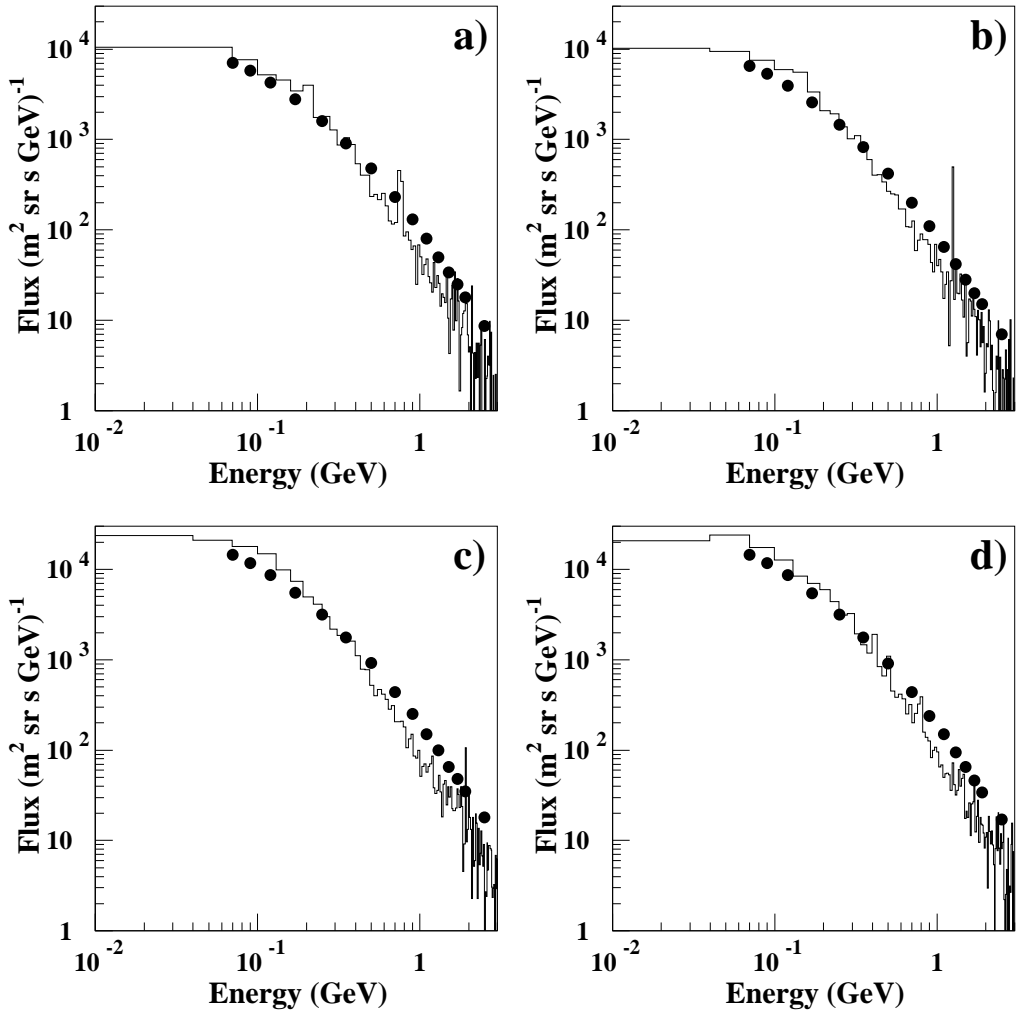,height=15 cm,width=16.3cm}}
\caption{%
Differential energy spectra of atmospheric neutrino (averaged over all directions)
at the Gran Sasso site. 
a) $\nu_{e}$, b) $\tilde{\nu}_{e}$ c) $\nu_{\mu}$, d) $\tilde{\nu}_{\mu}$. 
Dots from Ref.\cite{bib-Barr}. Histograms - this work.}
\label{fig:519a}
\end{center}
\end{figure}

\newpage
\begin{figure}[hp]
\begin{center}\mbox{
\epsfig{file=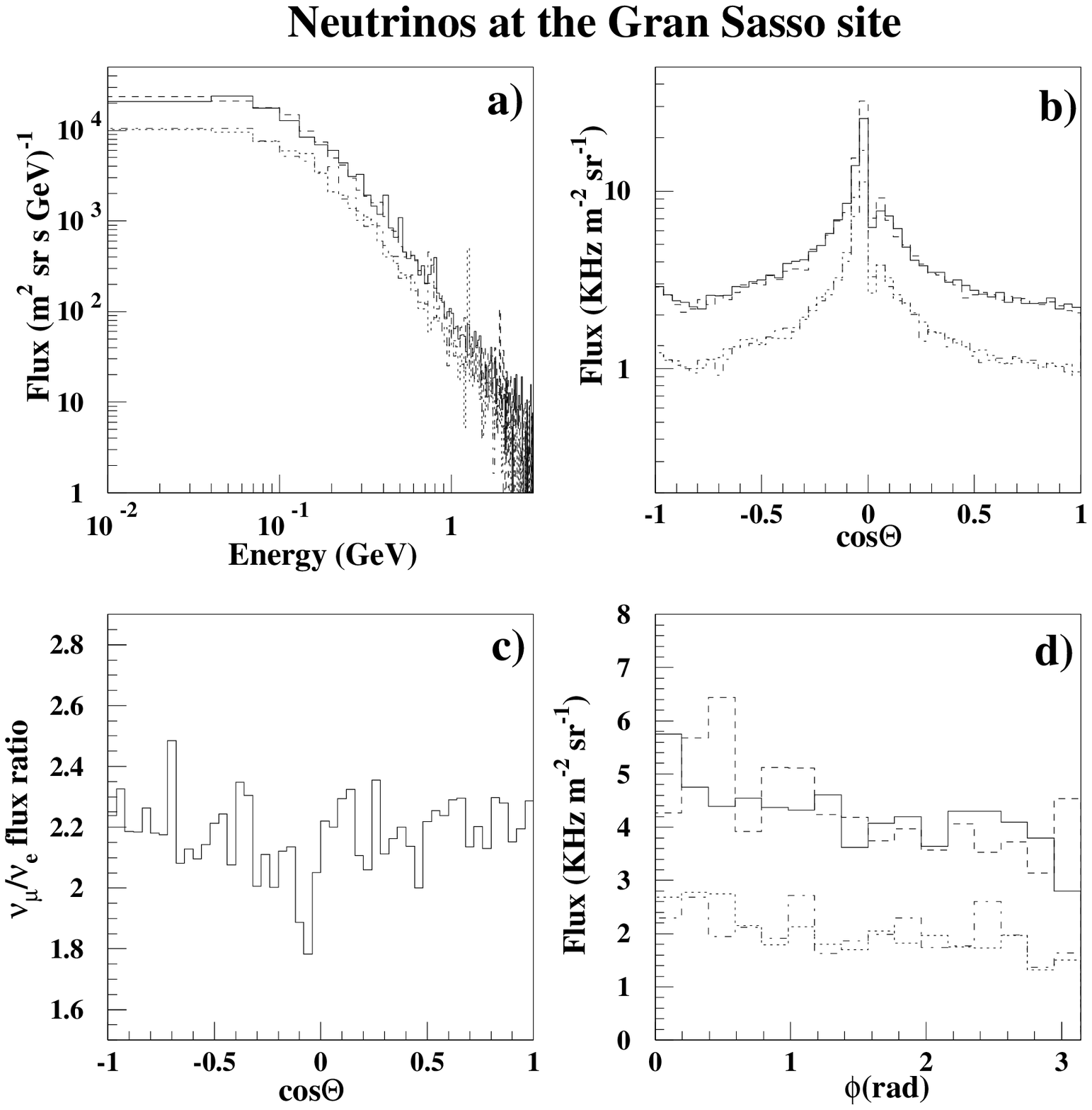,height=15 cm,width=16.3cm}}
\caption{%
Neutrino flux at the Gran Sasso site. 
a) Energy spectra. b) Nadir angle distribution. c) Flux ratio of $\mu/e$ neutrinos.
d) Azimuth angle distribution.
In a),b) and d) solid line - $\tilde{\nu}_{\mu}$, dashed line - $\nu_{\mu}$,
dotted line - $\tilde{\nu}_{e}$, 
dashed-dotted line - $\nu_{e}$.} 
\label{fig:519}
\end{center}
\end{figure}

\newpage
\begin{figure}[hp]
\begin{center}\mbox{
\epsfig{file=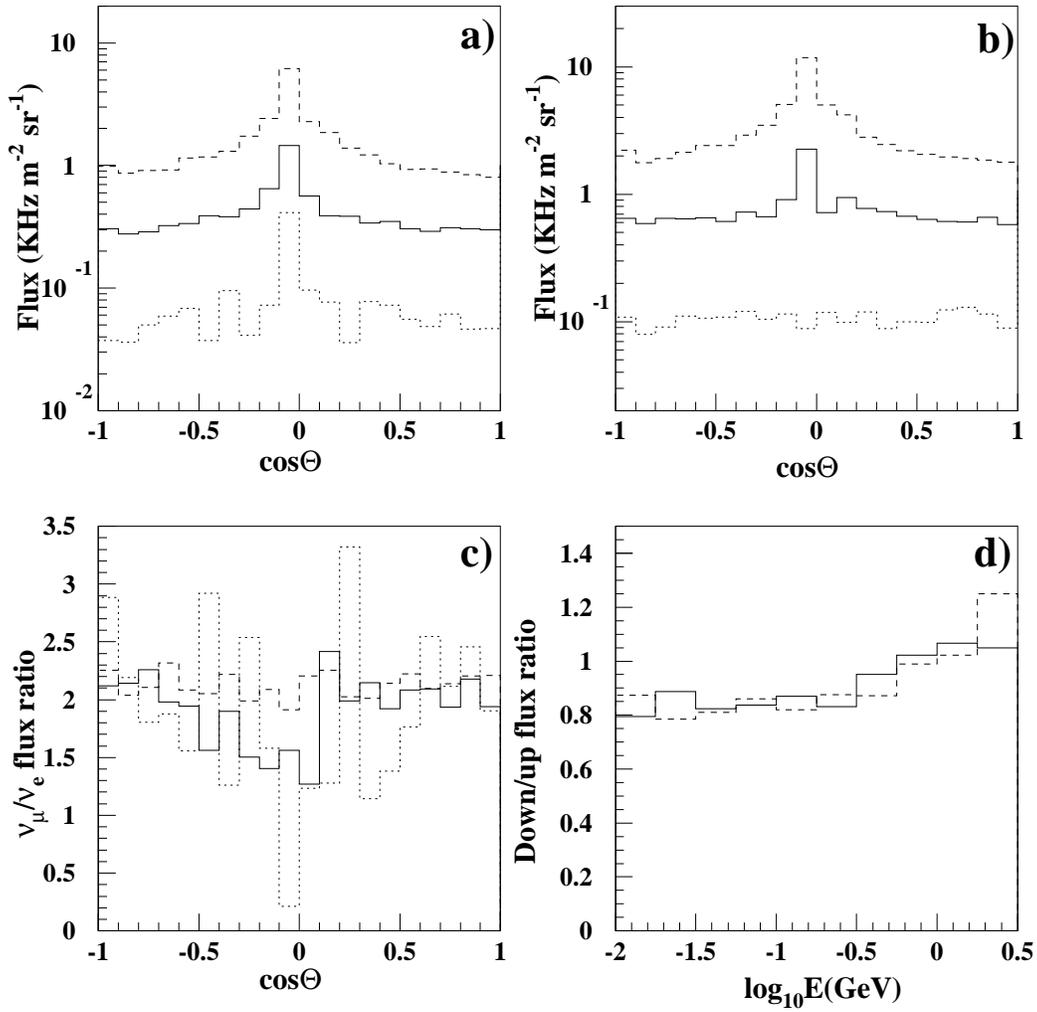,height=15 cm,width=16.3cm}}
\caption{%
Neutrino flux at the Gran Sasso site.
Nadir angle distribution for a) electron ($\nu_{e},\tilde{\nu}_{e}$) and b)
muon ($\nu_{\mu},\tilde{\nu}_{\mu}$) neutrinos of different energy.
d) Energy dependence of the ratio of 
the muon/electron fluxes.In a),b) and c) - dashed line 0.1$<E<$0.31 GeV, solid line 0.31$<E<$1.0 GeV,
dotted line $E>$1.0 GeV.
d) Energy dependence of the flux ratio of downgoing to upgoing muon (solid line) and
electron (dashed line) neutrinos.}
\label{fig:520}
\end{center}
\end{figure}

\newpage

This systematic effect for neutrinos with the highest flux (see Fig.\ref{fig:515}b) if not taken into 
account may lead to a conclusion about $\nu_{\mu},\tilde{\nu}_{\mu}$ disappearance. 
The east-west asymmetry of electron and muon neutrino is visible in Fig.\ref{fig:515}d. 

The energy dependence of the angular distribution of the flux of electron and muon neutrino is 
illustrated by Figs.\ref{fig:516}a and \ref{fig:516}b correspondingly. Note the absence of the
increase in the muon neutrino flux in the horizontal direction. 
Figure \ref{fig:516}d shows the energy dependence of the downward/upward flux ratio for muon (solid line)
and electron (dashed line) neutrinos. The ratio is the same for two different kinds of neutrino in the
energy range below 0.31 and above 1.5 GeV. However, for the region between 0.31 and 1.5 GeV where
the bulk of the experimental data is collected the down/up ratio for muon neutrinos is systematically
higher than that for electron neutrinos. This systematic effect may fake muon neutrino disappearance. 

Neutrino fluxes at the Soudan and Gran Sasso sites are calculated for the areas with geographical 
coordinates of $48^o\pm12^o~ N,~98^o\pm12^o~ W$ and $42^o\pm12^o~ N,~13^o\pm12^o~ E$ and the depth
underground of 700 m and 750 m correspondingly. 
As for the Kamioka site the averaged over all directions differential energy spectra of different
neutrinos and antineutrinos at the Soudan (Fig.\ref{fig:517a}) and Gran Sasso (Fig.\ref{fig:519a}) 
sites obtained in the present calculation give lower neutrino rates at higher energies than those 
obtained by other authors. The angular distributions of neutrino flux is strongly up/down asymmetric
with predominance of the downward flux at the Soudan site (Fig.\ref{fig:517}b) whereas it is almost
up/down symmetric for Gran Sasso (Fig.\ref{fig:519}b). The value and the energy dependence of the 
downward/upward flux ratio (Figs.\ref{fig:518}d,\ref{fig:520}d) for muon (solid line) and electron 
(dashed line) neutrinos is different for the two sites. At both experimental sites this ratio is 
almost the same for neutrinos of different kind with energies less than 1.5 GeV.  
There is no strong indication that the ratio of muon to electron fluxes at the Soudan site 
(Fig.\ref{fig:517}c,\ref{fig:518}d) depends on the neutrino direction. On the contrary the
tendency for the ratio of muon to electron fluxes to decrease for directions close to a horizontal 
plane is clearly seen for neutrinos in the energy range 0.31 - 1.0 GeV in Gran Sasso (solid line in 
Fig.\ref{fig:520}c).

\section{\bf Summary and conclusions}

The proton flux near the Earth consists of primary protons from space, primary protons scattered in the 
atmosphere and secondary protons produced in atmospheric showers. Each component of the proton flux 
has different angular distribution.
The results of the above Monte Carlo study of propagation in space and interaction with the atmosphere
of cosmic particles show that the primary proton flux in the vicinity of the Earth is essentially anisotropic 
at the directions approaching the horizontal plane. This result is in contradiction with the assumption 
of most atmospheric neutrino flux calculations. The anisotropy has an important impact on the production
of higher energy neutrino.  
 
The momentum spectra of primary protons from space display at different latitudes the 
expected \cite{bib-Stor,bib-Ster,bib-Sing} behaviour connected with the Earth's magnetic field (cutoff).
There exist, however, a non negligible amount of primary protons with momenta well below the magnetic cutoff.   

The momenta of the considerable number of scattered primary protons present at all magnetic latitudes 
are close to the cutoff value. When backtraced, the scattered protons have "forbidden" trajectories
and consequently their contribution to atmospheric neutrino production was not taken into account
in previous calculations of neutrino fluxes.

Predominantly secondary under cutoff particles (protons,$e^+,e^-$) are present in equal amount and
the same momenta in the fluxes of particles going to and from the Earth. The simulation correctly
describes the fluxes of protons and leptons ($e^+,e^-$) measured by AMS 
\cite{bib-AMS1,bib-AMS2,bib-AMS4}.

The atmospheric neutrinos produced in the simulation have the energy spectra, relative fluxes and
angular distributions different from those obtained in the previous atmospheric neutrino flux 
calculations 
\cite{bib-Hon1,bib-Hon,bib-Barr,bib-Barr1,{bib-Vol},{bib-Bug},{bib-But},bib-Batt,bib-LipGeo}. 
The difference is a direct consequence of the anisotropy of the cosmic ray flux at the
vicinity of the Earth. As it follows from the present study, the flux of primary cosmic 
particles at their entrance into the atmosphere is diminishing when approaching the horizontal
direction (see Fig.\ref{fig:505o}). Because of that, the fraction of high energy secondaries capable 
to decay before they lose most of their energy on the ionization of the atmosphere decreases. 
This results in smaller neutrino fluxes at higher energies.

The up/down asymmetry of the neutrino fluxes resulting from this study is essentially different
from the previous calculations. Similar to previous 3-D calculations the flux of atmospheric
neutrinos has a maximum in the angular region around the horizontal plane. However, as it is 
demonstrated in this study there is an important contribution to the horizontal flux from neutrinos 
produced just under the Earth's surface. A part of the flux of the "underground" neutrinos can 
be detected in the underground experiments. The horizontal region is the most important for 
determination of the up/down asymmetry of the atmospheric neutrino fluxes and consequently the 
result of the present study may affect the eventual experimental measurement of the neutrino 
oscillation parameter $\rm \Delta m^2$. Calculated for different experimental sites the angular 
distributions of neutrino fluxes of different kind and their dependence on the neutrino energy      
suggest the presence of effects faking disappearance of upward going muon neutrinos. These 
effects are different at different experimental sites (Figures \ref{fig:516}c,d;
\ref{fig:518}c,d,\ref{fig:520}c,d). This could probably be the reason why different experiments do not
agree in their estimate of oscillation parameters.     

{\bf Acknowlegements.} I'm grateful to Prof.Yu.Galaktionov for very useful and stimulating 
discussions. I have to express my gratitude to Dr.V.Shoutko for reading the manuscript and giving
suggestions for important improvements. Finaly I would like to thank Prof.S.C.C.Ting for his 
interest in this work and encouragement.  

\newpage

\newpage    


\end{document}